\documentclass[12pt,a4paper]{paper}%journal}
\usepackage[left=2.5cm,right=2.5cm,top=2.5cm,bottom=2.5cm]{geometry}

\usepackage{graphicx}
\usepackage{amsmath}
\usepackage{amsfonts}
\usepackage{amssymb}

\usepackage[latin1]{inputenc}

\newcommand{\dd}{\mathrm{d}}
\newcommand{\vv}[1]{\overrightarrow{#1}}

\graphicspath{{./figs/}}
 \title{Impact of shape of container on natural convection and melting inside enclosures used for passive cooling of electronic devices }
 
\author{K. El Omari$^1$ \and T.  Kousksou \and Y.  Le Guer}                                                                                                         
\institution{Laboratoire des Sciences de l'Ingénieur Appliquées à la Mécanique\and et au Génie Electrique (SIAME)\and 
 Université de Pau et des Pays de l'Adour (UPPA)\and 
 Bat. D'Alembert, rue Jules Ferry, BP. 7511, 64075 Pau cedex - France}

\begin{document}
\maketitle
\footnotetext[1]{Corresponding author: kamal.elomari@univ-pau.fr\\ Preprint published in Applied Thermal Engineering, Volume 31, Issue 14-15, October 2011, Pages 3022-3035.}

%%%%%%%%%% Elseveir %%%%%%%%%%%%%%%
%\begin{frontmatter}
% \title{\textbf{\Large Impact of shape of container on natural convection and melting inside enclosures used for passive cooling of electronic devices}}
% \author[ky]{Kamal El Omari\corref{cor1}}
% \ead{kamal.elomari@univ-pau.fr}
% \author[ky]{Tarik Kousksou}
% \ead{tarik.kousksou@univ-pau.fr}
% \author[ky]{Yves Le Guer}
% \ead{yves.leguer@univ-pau.fr}
% 
% \address[ky] {Laboratoire des Sciences de l'Ingénieur Appliquées à la Mécanique et au Génie Electrique (SIAME)\
% {F\'ed\'eration IPRA-CNRS, Université de Pau et des Pays de l'Adour (UPPA)}\\
% {Bat. D'Alembert, rue Jules Ferry, BP. 7511, 64075 Pau cedex - France}}
% 
% \cortext[cor1]{Corresponding author, Tel: +33 559 407 147, Fax: +33 559 407 160}
%%%%%%%%%% Elseveir %%%%%%%%%%%%%%%

 \begin{abstract}
 The present paper numerically analyzes a passive cooling system using enclosures with different geometries filled with thermal conductivity-enhanced phase change material (PCM). A numerical code is developed using an unstructured finite-volume method and an enthalpy-porosity technique to solve for natural convection coupled to a solid-liquid phase change. Five geometries containing the same volume of PCM are compared while cooling the same surface. The unsteady evolution of the melting front and the velocity and temperature fields is detailed. Other indicators of cooling efficiency are monitored, including the maximum temperature reached at the cooled surface. The computational results show the high impact of varying geometry: a maximum temperature difference as high as 40 °C is observed between two of the enclosures. The best efficiency is obtained for an enclosure shifted vertically relative to the cooled surface. Other findings and recommendations are made for the design of PCM-filled enclosures.
\end{abstract}

%\begin{keyword}
%Passive cooling \sep  phase change material \sep   melting \sep   natural convection \sep   enthalpy-porosity method \sep   unstructured finite-volume method
%\end{keyword}
%\end{frontmatter}


\section{Introduction}

The melting of phase change materials (PCM) coupled to natural convection in enclosures has been studied extensively. This situation is encountered in many technical applications, such as latent heat storage systems \cite{elomari2004,kousksou2005} and thermal insulation for buildings \cite{alawadhi2008,Ahmad2006}. The melting of PCM is also used to control the temperature of the surface of electronic components that release instantaneous or periodic high density heat fluxes to moderate the need for classical cooling devices. There has been increasing interest in this type of passive cooling for electronic circuits, such as chipsets, laptop processors or graphics cards \cite {Kandasamy2008, nayak2006}. These elements are continuously becoming smaller, and their released heat densities are increasing; therefore, they require efficient, economic, and silent cooling systems. A particular application for such systems is the cooling of electrical devices inside shelters located in difficult-to-access sites and subject to periodic temperature changes. In these situations, cooling solutions that do not use moving mechanical elements, such as fans, are more suitable to avoid the need for frequent maintenance.
 
Several authors have studied phase changes and natural-convection-dominated melting in enclosures. Beyond the development and application of numerical models and experimental techniques, these authors have proposed several methods and concepts to enhance the heat transfer and melting rate. Gong et al. \cite{gong1999} demonstrated the positive effect of inverting a square PCM container heated from the side during the melting process when thermal stratification in the melt slows the melting rate. For the same geometry, Khodadadi and Hosseinizadeh \cite{khodadadi2007} numerically investigated the effect of adding nanoparticles to the PCM. They showed that when the enclosure is cooled laterally, the rate of freezing increases because of the presence of nanoparticles. Starting from a study of PCM melting in a square enclosure heated from below, Fteïti and Ben Nasrallah \cite{fteiti2005} examined the impact of the aspect ratio of the PCM-filled  enclosure and found that flat enclosures exhibit faster melting but have a lower asymptotic limit of the total melt ratio  (they imposed a temperature boundary condition equal to the melting temperature  at the top side of the enclosure and a lower temperature at the bottom). Hernandez-Guerrero et al. \cite{hernandez2005} also studied the impact of the aspect ratio with a different numerical model and found similar qualitative results for the case of tall enclosures. It is worth noting that the two latter works compared differently shaped enclosures that contained varying quantities of PCM.

Several researchers have considered the cooling of a vertical surface that dissipates heat flux using a tall enclosure filled with PCM. Binet and Lacroix \cite{binet2000} performed numerical and experimental studies to analyze the impact of the positions of three heat sources, their sizes and the aspect ratio of the enclosure itself. A similar numerical study was conducted by Krishnan and Garimella \cite{krishnan2004} on the conjugate heat transfer through the enclosure walls. Pal and Joshi \cite{pal2001} studied a uniformly dissipating heat source using experiments and numerical modeling, and they established correlations for the temporal evolution of the parietal heat transfer and the melt fraction. Huang et al. \cite{huang2006} used a three dimensional (3D) numerical model to investigate the cooling of photovoltaic cells by a PCM enclosure equipped with metallic internal fins.

A large amount of research has been conducted to investigate the applicability of cooling by PCM for electronic and electric devices. For example, the work in \cite{tan2004} focused on the cooling of a mobile device using embedded PCM enclosures, \cite{shatikian2008, akhilesh2005, wang2008, nayak2006} focused on PCM-based heat sinks, while in \cite{kizilel2009}  PCM was used to control the temperature of Li-ion batteries.

Unlike for thermal insulation, the efficiency of passive cooling with PCM is closely related to the rapidity of its melting: the temperature of the heat-dissipating surface does not rise rapidly because of the absorption of the latent heat of fusion. 

The speed with which the PCM melts in a laterally heated enclosure is controlled by the conductivity of the solid PCM during the early stages and by the natural convection currents in the melt during the later stages. This natural convection flow depends on the thermo-physical properties of the PCM and the geometry of the enclosure. Thus, the shape and extent of the enclosure containing the PCM have an important effect on the kinetics of the melting front and, therefore, on the temperature of the cooled wall.

In this work, we are interested in the cooling of a vertical surface releasing a fixed heat flux by means of a PCM-filled enclosure that entirely covers the surface. The aim of this study is to propose and examine different geometric shapes and relative positions of the enclosure. Numerical simulation is a well-suited tool for this purpose, especially in the early stages of design. It allows different test situations to be compared before an experimental prototype is constructed. A numerical model based on a fixed grid method and applied to an unstructured finite-volume formulation has been developed for this purpose. 

In the following sections, the studied system is described with the different considered geometries and the flow conditions are then detailed. Next, the physical model and the numerical method are presented, and the implemented code is validated. Afterwards, detailed results are presented concerning the evolution of the velocity and the temperature fields, the melting front and the rate of fusion. Finally, the geometric impact on the coupled effects of natural convection and phase change is analyzed. Conclusions are then drawn based on these results.

\section{The studied system}

In this study, we are interested in the case of a vertical wall releasing a uniform heat flux at a high rate but for a limited duration. We aim to maintain this surface (designated $S_h$) at a sufficiently low temperature to avoid damaging the electrical component behind it. To achieve this objective, we provide this surface with an enclosure filled with a PCM. The surface to be cooled constitutes one face, or a part of a face, of this enclosure while the other faces are exposed to the external air and are subject to natural convection. When the heat release from the surface $S_h$ begins, its temperature rises until it reaches the fusion temperature of the PCM, which causes the material to melt. Natural convection currents take place progressively in the melt, and the generated flow ensures that heat is transferred from the surface $S_h$ to the solid PCM and maintains the melting process. The strength of the fluid flow involved in this process is closely related to the shape of the formed melted area, the extent of which is limited by the boundaries of the enclosure. Therefore, we expect that the efficiency of the heat transfer and the rate at which PCM melts, strongly depend on the shape of the enclosure. This simple passive cooling device is appropriate for intermittent heat release and is studied here for a uniform heat flux rate of $10^4\ W.m^{-2}$. For continuous heat release and higher heat flux densities, more sophisticated solutions could be considered such as heat pipes \cite{reay2007}.

The ``natural'' or ``intuitive'' choice of the shape of the PCM container can be regarded as the rectangular geometry which has been extensively studied in the literature for its simplicity. It can also be considered that the circular shape has the advantage of the absence of corners that may retain non melted PCM. Furthermore, the ``intuitive'' choice of the relative position of the container to the heating surface would be a centered position. Thus, in this study we aim to verify the relevance of these different choices. Consequently, we proceeded to investigate the impact of different enclosure shapes on the natural convection flow coupled to the solid-liquid phase change. The five different geometries studied contained the same volume of PCM. Fig. \ref{fig:meshes}  shows the computational meshes used to model the enclosures (the chosen meshes densities are described in section \ref{sec:mesh}). In this figure, the location of the surface $S_h$ is indicated in gray. The area of $S_h$ is the same for all of the enclosures. These geometries (detailed in the next section) were not all considered at the beginning of this work but were designed progressively given the results obtained from the first tested shapes. However, the results obtained for the five enclosures will be presented together. Some of the enclosures are rectangular, whereas others have rounded corners. Barlett et al. \cite{barletta2006} have studied natural convection in both rectangular and rounded-corner enclosures and have shown that the latter shape substantially improves heat transfer. Thus, we considered rounded shapes in the present study and investigated their potential role in situations involving a phase change.

A known drawback of the considered cooling system is the relatively low thermal conductivity of common PCMs (e.g., wax paraffin), which may cause a high temperature peak at the surface in the case of high heat flux early-on in the heat release, when the heat transfer occurs only by conduction. To overcome this limitation, several solutions have been investigated in the literature, such as utilizing fins \cite{nayak2006}, including a high thermal conductivity metallic foam \cite{nayak2006, mesalhy2005} or adding nanoparticles. Extensive reviews of all these solutions can be found in \cite{jegadheeswaran2009, agyenim2010}.
 In our study, we consider a solution that includes graphite nanoparticles with high thermal conductivity in the PCM. These types of materials have experienced extensive development and exhibit high potential for use in several applications. Many researchers \cite{kim2009,sari2007} have demonstrated that the thermal conductivity of these products can reach unity (SI units). As indicated above, in the present study, we focus on the impact of the enclosure shapes on the melting kinematics, natural convection strength and temperature extremes at the cooled surface. 
Therefore, we choose not to consider the details of the mixture of PCM and nanoparticles and the relationship between the percentage of nanoparticles and the thermophysical properties of the obtained PCM, such as viscosity, effective conductivity and specific heat capacity. Instead we consider a model PCM with constant thermophysical properties but enhanced thermal conductivity (see next section for details).

\section{Flow conditions and PCM properties}\label{sec:FL}

\begin{figure}[tbp]
  \begin{center}
\includegraphics[width=0.75\linewidth]{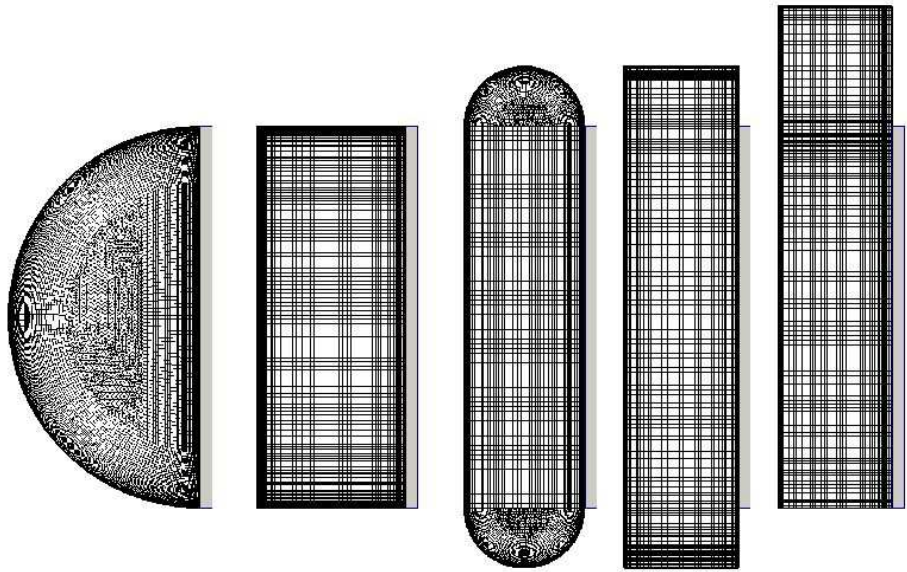} 
\caption{\it \small The geometries of the five studied enclosures: \textbf{a},\textbf{b}, \textbf{c}, \textbf{d}, and \textbf{e} (from left to right) and corresponding meshes (in gray: the position of the surface $S_h$).}\label{fig:meshes}
  \end{center}
\end{figure}

As stated above, in this work, we study the melting of PCM inside five enclosures of different shapes that surround the same volume. We limit our investigation to two-dimensional configurations that can be modeled in a vertical plane. One of the sides of the first two enclosures on the left in Fig \ref{fig:meshes} (\textbf{a} and \textbf{b}) exactly matches the surface $S_h$. The first enclosure is rounded (a half disc), and the second is rectangular. The three other enclosures (\textbf{c}, \textbf{d} and \textbf{e}) have sides that vertically exceeded the surface $S_h$, and thus, their widths are smaller than those of enclosures (\textbf{a}) and (\textbf{b}). Enclosure (\textbf{c}) is oblong with rounded corners, whereas enclosure (\textbf{d}) is rectangular. The last enclosure (\textbf{e}) has the same geometry as enclosure (\textbf{d}), but it is translated upward such that its left and right sides exceed the surface $S_h$ only at the top. The surface $S_h$ has a fixed height of $H=5\ cm$, whereas the dimensions of the five enclosures are adjusted to contain the same volume. These dimensions are given in Tab. \ref{tab:cavities} in terms of maximum width and height.

 \begin{table}[tbh]
 \caption{Maximum widths $w_{max}$ and heights $h_{max}$ of the five enclosures (in cm).\label{tab:cavities}}
 \centering
\begin{tabular}{c c c}
\hline
Enclosure & $w_{max}$ & $h_{max}$\\
\hline
(\textbf{a}) & 2.5000 & 5.0000\\
(\textbf{b}) & 1.9635  & 5.0000\\
(\textbf{c}) & 1.5742 & 6.5742\\
(\textbf{d}) & 1.4933 & 6.5742\\
(\textbf{e}) & 1.4933 &  6.5742\\
\hline
 \end{tabular}  
\end{table}

The rate of heat flux at the surface $S_h$ is fixed to a constant value, $q'' = 10^4\ W.m^{-2}$, for all cases. The heat transfer to the external air at the other boundaries is considered to occur by natural convection and is modeled by a heat transfer coefficient $h = 30 \ W.K^{-1}.m^{-2}$, whereas the external air temperature is assumed to be equal to the initial temperature of the PCM, $T_\infty = T_0=20$ °C.
 
We consider a model PCM in this study composed of a pure paraffin wax and graphite nanoparticles. To simplify the analysis in the study, the properties of this PCM are considered to be identical in the solid and liquid phases and non-temperature dependent (except in the body forces term). The values of the adopted properties are listed in Tab. \ref{tab:MCP}. A maximum temperature difference $\Delta T$ characterizing the flow can be obtained by considering the final or maximum admissible temperature that reaches the surface $S_h$, which corresponds to the temperature at which an electronic device is damaged (i.e., $90$ °C):  $\Delta T = 90 - 20 =70$  °C.  The Rayleigh number based on this temperature difference and on the height $H$ of the surface $S_h$ is $Ra = 2.7\times 10^7$, which corresponds to a laminar flow regime. The Stefan number corresponding to this limit situation is $Ste = 0.87$.
 
\begin{table}[tbh]
\caption{Thermophysical properties of the model enhanced-conductivity PCM \label{tab:MCP}.}
\centering
\begin{tabular}{l l}
\hline
Dynamic viscosity ($\mu$)& $5\times 10^{-3} Pa.s$\\
Density ($\rho$) & 800 $kg.m^{-3}$\\
Thermal conductivity ($k$) & 1 $W. m^{-1} K^{-1}$\\
Specific heat ($C_{p}$) & 2500 $J. kg^{-1} K^{-1}$\\
Latent heat ($L$) & $200\ kJ.kg^{-1}$\\
Thermal dilatation coefficient ($\beta$) & $10^{-3}\ K^{-1}$\\
Prandtl's number ($Pr$) & $12.5$\\
Melting temperature ($T_m$) & $20$ °C\\
\hline
 \end{tabular}  
\end{table}

\section{Computational modeling}
\subsection{The governing equations}\label{goveqnummeth}

The unsteady equations governing the flow of incompressible non-isothermal fluids are solved over the entire computational domain. The three-dimensional equations of conservation of momentum, mass and energy (in terms of temperature) are considered valid for both solid and liquid phases, which are distinguished by a liquid volume fraction $f$ that takes values 0 and 1, for solid and liquid phases, respectively. These conservation equations are considered in their integral forms:
\begin{eqnarray}
 \dfrac{\partial}{\partial t} \int_{V}\rho\ \vec U\ \dd V
   + \int_{S} \rho\ \vec U \vec U \cdot \vec n \ \dd S& =&
   - \int_{V} \vec\nabla p \ \dd V  
+ \int_{S} \overset{=}{\tau}\cdot \vec n \ \dd S + \int_{V} S_U \ \dd V  \label{eq:mvt}\\
 \int_{S} \vec U \cdot \vec n \ \dd S &=&0 \label{eq:4}\\
\dfrac{\partial}{\partial t} \int_{V}\rho\ C_p T\ \dd V
   + \int_{S} \rho\ C_p T\ \vec U \cdot \vec n \ \dd S &=&
   \int_{S} k\ \vec\nabla T\cdot \vec n \ \dd S + \int_{V} \rho L\dfrac{\partial f}{\partial t} \dd V  
\label{eq:energy}
\end{eqnarray}
where $\vec U$ is the velocity vector, $p$ the pressure and $T$ the temperature. $\overset{=}{\tau}$, is the viscous stress tensor for a Newtonian fluid:
\begin{equation}
    \overset{=}{\tau}= \mu \left( \overset{=}{\nabla}U+(\overset{=}{\nabla}U)^T \right)
\end{equation}

The integration occurs over a volume $V$ surrounded by a surface $S$, which is oriented by an outward unit normal vector $\vec n$. The source term in the momentum conservation equation (Eq. (\ref{eq:mvt})) contains two parts:
\begin{equation}
S_U = \rho \beta (T-T_{ref}) \vec g + A\ \vec U \label{eq:source}
\end{equation}
where $\beta$ is the coefficient of volumetric thermal expansion and $\vec g$ the acceleration of gravity vector. The first part of this source term represents the buoyancy forces due the thermal dilatation. For sake of simplicity, the Boussinesq's approximation is used in this comparative study that focuses on the impact of different geometries. The $T_{ref}$ temperature was chosen as the mean temperature of the PCM liquid phase and was recalculated at each time step. Therefore, $T_{ref}$ is more representative of the temperatures in the liquid phase throughout the whole process, especially when all of the PCM has melted. The second part of the source term (Eq. \ref{eq:source}) is a penalization term that ensures zero velocity in the computational cells where the PCM is solid, i.e., where $f=0$. The penalization coefficient $A$, based on the Carman-Koseny relation for a porous medium, is written as a function of $f$ as follows:
\begin{equation} A = -\dfrac{C(1-f)^2}{f^3+\epsilon},\end{equation}
where $C=1.6\times 10^6$, and $\epsilon=10^{-3}$. This coefficient ensures a smooth transition between solid and liquid media. In the case of pure PCM, the phase change front is sharp, and the transition takes place over only one computational cell in the direction of the front displacement.

In Eq. (\ref{eq:energy}), the last term of the right hand side (RHS) introduces the effect of the latent heat of phase change $L$ into the energy equation. This effect is accounted for by considering the variation in time of the liquid fraction $f$. 

This general three-dimensional (3D) model can be restricted to deal with 2D computations (e.g., in the ($\vec x,\vec y$) plan) without any change by considering a single layer of computational cells (in the $\vec z$ direction) and by neutralizing the top and bottom faces (with respect to $\vec z$). It is this approach that was used for all the computations presented in this study, thus, the considered meshes are composed of one layer of hexahedra with or without prisms.
 
\subsection{The numerical method}

The conservation equations (\ref{eq:mvt}, \ref{eq:4} and \ref{eq:energy}) are solved by implementing them in an in-house code called Tamaris. This code has a three-dimensional unstructured finite-volume framework that is applied to hybrid meshes. The variable values ($\vec U$, $p$, $T$ and $f$) are stored in cell centers in a collocated arrangement. The cell shapes can vary (e.g., tetrahedral, hexahedral, prismatic or pyramidal). 

\begin{figure}[tb]
 \centering
 \begin{center}
 \includegraphics[width=0.5\textwidth]{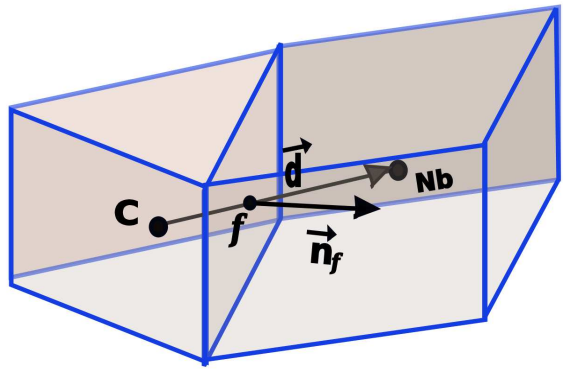}
\end{center}
 \caption{A computational cell $C$ and one of its neighbors $N_b$.}
 \label{fig:cell}
\end{figure}

To describe the discretization method used in the code, we can write Eqs. (\ref{eq:mvt}) and (\ref{eq:energy}) in the generic convection-diffusion form with respect to a conserved variable $\phi$:
\begin{equation}
 \dfrac{\partial}{\partial t} \int_{V}\rho \phi\ \dd V 
    + \int_{S} \rho \phi \vec U \cdot \vec n \ \dd S = 
    \int_{S} \Gamma \vec\nabla \phi\cdot \vec n \ \dd S + \int_{V} S_\phi\ \dd V,
\label{eq:generic}\end{equation}
where $\Gamma$ is a diffusion coefficient and $S_\phi$ a source term. The spatial schemes used to approximate the diffusive and convective fluxes are both second-order accurate. The diffusion term is discretized by approximating the surface integrals with a sum over all cell faces $f$ (Fig. \ref{fig:cell}):
\begin{equation}
 \int_{S} \Gamma \vec\nabla \phi\cdot \vec n \ \dd S = \sum_f \Gamma_f A_f (\vv{\nabla\phi})_f \cdot \vec n_f, 
\end{equation}
where $A_f$ is the area of face $f$. For unstructured meshes, orthogonality is an exception, and it needs to be handled correctly. Thus, the normal gradient $(\vv{\nabla\phi})_f \cdot \vec n_f$ is decomposed into an implicit contribution that uses the values of $\phi$ at the centers of the two cells sharing the face $f$ (the first term on the RHS of Eq. (\ref{eq:diffu})) and a non-orthogonality correction term treated explicitly by a deferred approach to preserve the second-order accuracy of the centered differencing. We use the over-relaxed decomposition suggested by \cite{jasak1996} to enhance the convergence properties of the discretization of the diffusive term:
\begin{equation}
 (\vv{\nabla\phi})_f \cdot \vec n_f = \dfrac{\phi_{N_b} - \phi_c}{||\vec d||}  \dfrac{1}{\vec d \cdot \vec n_f}+ \overline{\vv{ \nabla \phi}}
\cdot \left(\vec n_f - \frac{\vec d}{\vec d \cdot \vec n_f }\right)
\label{eq:diffu}
\end{equation}
$\vec d$ is the vector joining the centers of the two adjacent cells (see Fig. \ref{fig:cell}). The average gradient $\overline{\vv{ \nabla \phi}}$ is interpolated from the gradients of these neighboring cells. 

The gradients of the variables at the cell centers are computed by Gauss' theorem:
\begin{equation}
\vv{ \nabla \phi} = \dfrac{1}{V} \int_{S}  \phi \ \vec n \ \dd S
= \dfrac{1}{V} \sum_f \phi_f A_f\ \vec n_f, \label{eq:gradgauss}
\end{equation}
where $\phi_f$ is the mean value of the variable interpolated using the values at the centers of two cells sharing face $f$:
\begin{equation}
\phi_f = \xi \phi_c + (1-\xi) \phi_{N_b}\quad \text{with}\quad \xi = \dfrac{\overline{fN_b}}{\overline{CN_b} }
\end{equation}
 Once the gradient is calculated for all computational cells, the values are used to determine a new estimate of $\phi_f$ as follows:
\begin{equation}
 \phi_f = \frac{1}{2} \left[ \left(\phi_{N_b} + \vv{ \nabla \phi}_{N_b} \cdot \vv{N_b f}\right) + \left(\phi_c + \vv{ \nabla \phi}_c \cdot \vv{C f} \right)\right]
\end{equation}
These new values of $\phi_f$ are used to re-compute the gradients more accurately using Eq. (\ref{eq:gradgauss}) \cite{ferziger2002}.

Convection terms are also transformed into a sum over faces $f$ by decomposing the surface $S$:
\begin{equation}
\int_{S} \rho \phi \vec U \cdot \vec n \ \dd S = \sum_f (\rho\phi A)_f \vec U_f\cdot \vec n_f,
\label{eq:8}
\end{equation}
where the face values $\phi_f$ require appropriate interpolation to be accurate and bounded. Thus, we use the non-linear high-resolution (HR) bounded scheme CUBISTA by Alves \textit{et al.} \cite{alves2003} in the $\gamma$ formulation of Ng \textit{et al.} \cite{ng2007}, where they expressed $\phi_f$ is a function of the upwind (UP) value of $\phi$ and its centered differencing (CD) value:
\begin{equation}
 \phi_f^{HR} = \phi^{UP} + \gamma(\phi_f^{CD}-\phi^{UP}).
\label{eq:HR}\end{equation}
The coefficient $\gamma$ is determined for each face based on the local shape of the flow solution using the normalized variable diagram (NVD) framework and observing the convection boundedness criterion (CBC) \cite{gask1988}. The first term of the RHS of Eq. (\ref{eq:HR}) is accounted for implicitly, whereas the second term is treated explicitly with the deferred-correction practice.

The pressure-velocity coupling is ensured by the SIMPLE algorithm \cite{patankar1980}, whereas the mass fluxes at the cell faces are evaluated using the Rhie-Chow interpolation \cite{rhie1983} to avoid pressure checkerboarding. The implicit three-time-step Gear's scheme \cite{gear1971} of second-order accuracy is used to discretize the unsteady terms:
\begin{equation}
 \dfrac{\partial}{\partial t} \int_{V}\rho \phi\ \dd V 
= \dfrac{3(\rho \phi)_c^n - 4(\rho \phi)_c^{n-1} + (\rho \phi)_c^{n-2} }{\Delta t} V
\end{equation}
The superscript $n$ stands for the current time step and $\Delta t = t^n-t^{n-1}$ is the time step. The RHS of Eq. (\ref{eq:generic}) is taken at time $t^n$. By applying the former discretizations, the generic conservation Eq. (\ref{eq:generic}) transforms into the algebraic form:
\begin{equation}
 a_c \phi_c +\sum_{N_b} a_{N_b} \phi_{N_b} = b_{c}. \label{eq:algeb}
\end{equation}

\begin{figure}[tbp]
\begin{center}
 \includegraphics[width=0.49\textwidth]{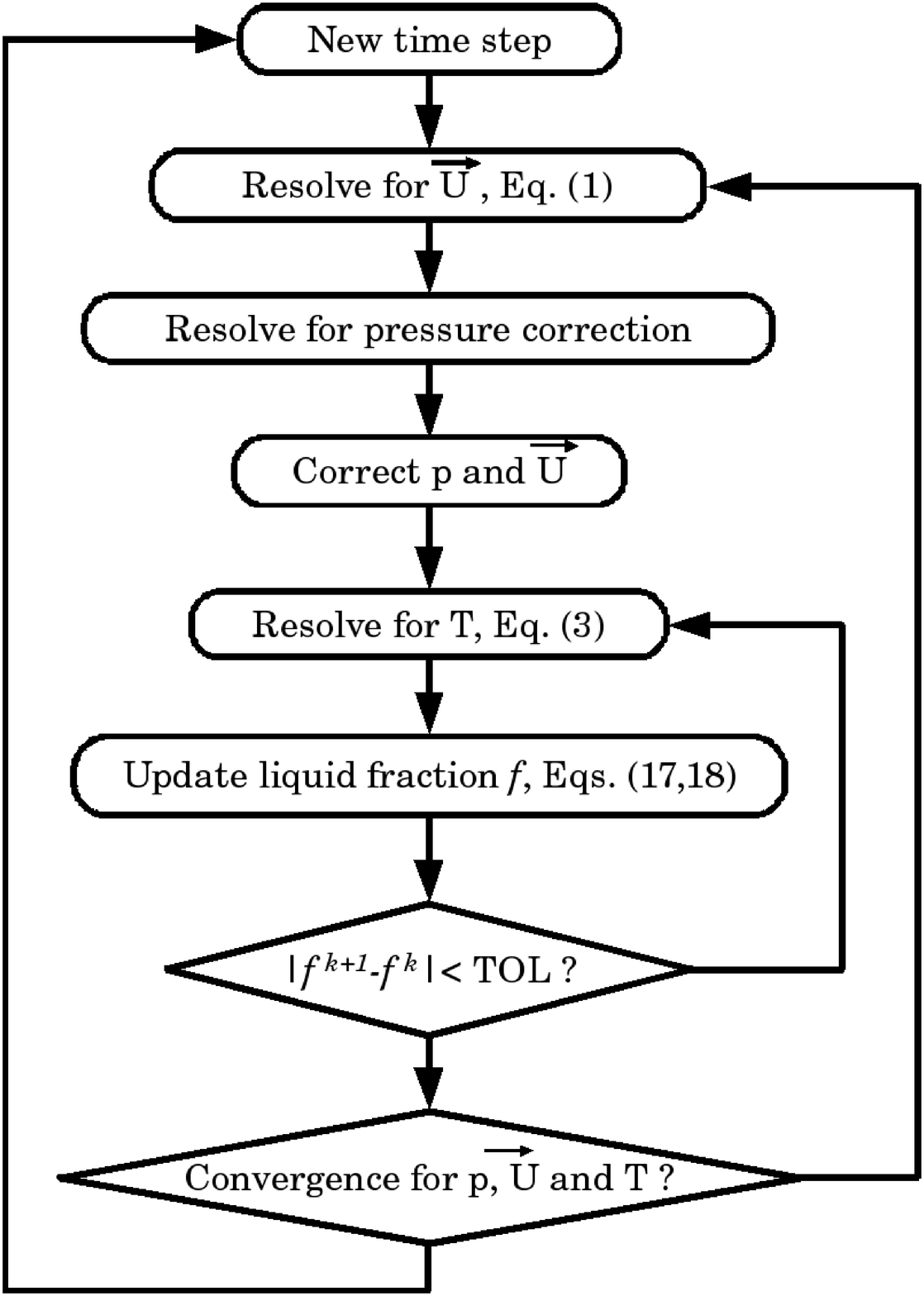} 
\caption{\it \small The liquid fraction updating procedure as included in the SIMPLE algorithm. \label{fig:algo}}
\end{center}
\end{figure}

Within each iteration of the SIMPLE algorithm, after the resolution of the momentum equation and of the Poisson equation for the pressure correction \cite{ferziger2002,patankar1980}, the energy equation is solved, and the fluid fraction $f$ is updated. These last two steps are repeated until the variation of $f$ is sufficiently small; then, the next SIMPLE iteration starts unless the convergence for $\vec U$, $p$ and $T$ is achieved, in which case a new time step is considered. A diagram of this algorithm is given in Fig. \ref{fig:algo}. The resolution of the energy equation is integrated in the SIMPLE iteration to take into account the high level of its coupling with the momentum equation through the body forces term. Additionally, the liquid fraction correction is iterated with the energy equation resolving to tightly couple the temperature field with the liquid fraction field.

In this study, the liquid fraction $f$ is updated by the ``\textit{new source}'' algorithm proposed by Voller \cite{voller90}, where the new value of $f$ at iteration $k+1$ and in cell $c$ is calculated as follows:
\begin{equation}
 f_c^{k+1} =  f_c^{k} + \dfrac{\Delta t\; a_c^k}{\rho L V} (T_c - T_m),
\end{equation}
where $T_m$ is the melting temperature of the PCM, and $a_c^k$ is the coefficient of $T_c$ in the discretized Eq. (\ref{eq:algeb}) that corresponds to temperature. This update is followed by an overshoot/undershoot correction:
\begin{equation}
f_c^{k+1} =
\begin{cases}
   0\; \text{ if } f_c^{k+1}<0\\
   1\;  \text{ if } f_c^{k+1}>1.
\end{cases}
\end{equation}

Following the ``\textit{new source}'' algorithm \cite{voller90}, the energy equation is penalized in the computational cells belonging to the phase change front ($0<f_c<1$) to ensure that the temperature at these cells is equal to the melting temperature $T_m$. This procedure is performed by adding a penalization source term, equal to $10^9\times T_m$, to the energy equations corresponding to these cells. This practice accelerates the convergence of the $f$ updating algorithm, and the number of iterations needed to reach convergence ($|f_c^{k+1} -  f_c^{k}|< TOL$) may be lowered to 1 or 2 depending on the size of the mesh and time-steps. The value of $TOL$ is fixed in this work to $10^{-4}$.  At each iteration, the discretization technique presented above leads to a linear system of algebraic equations in the form of Eq. (\ref{eq:algeb}) with a non-symmetric sparse matrix for each variable. These linear systems are solved using an ILU-preconditioned GMRES procedure implemented in the IML++ library \cite{dongarra1994}. In the scope of this work, all of the computational meshes were generated using the open-source software Gmsh \cite{geuzaine2009}.

\subsection{The code validation}

The present code has been successfully validated in some situations involving flow and heat transfer, as in \cite{elomari2009a, elomari2010b}. An additional code validation concerned with pure natural convection is given in \ref{sec:append}. We focus here on validating the code when applied to the case of melting of a pure PCM coupled to natural convection in the melt. The chosen test case is the 2D numerical benchmark presented by Hannoun \textit{et al.} \cite{hannoun2005}, which involves melting tin in a square enclosure subject on one side to a temperature higher than the melting temperature. The authors have presented extensive results obtained using a second-order accurate finite-volume method in structured meshes as fine as 600 $\times$ 600. 

\begin{figure}[tbp]
\begin{center}
% use packages: array
%\begin{tabular}[c]{cc}
 \includegraphics[width=0.49\textwidth]{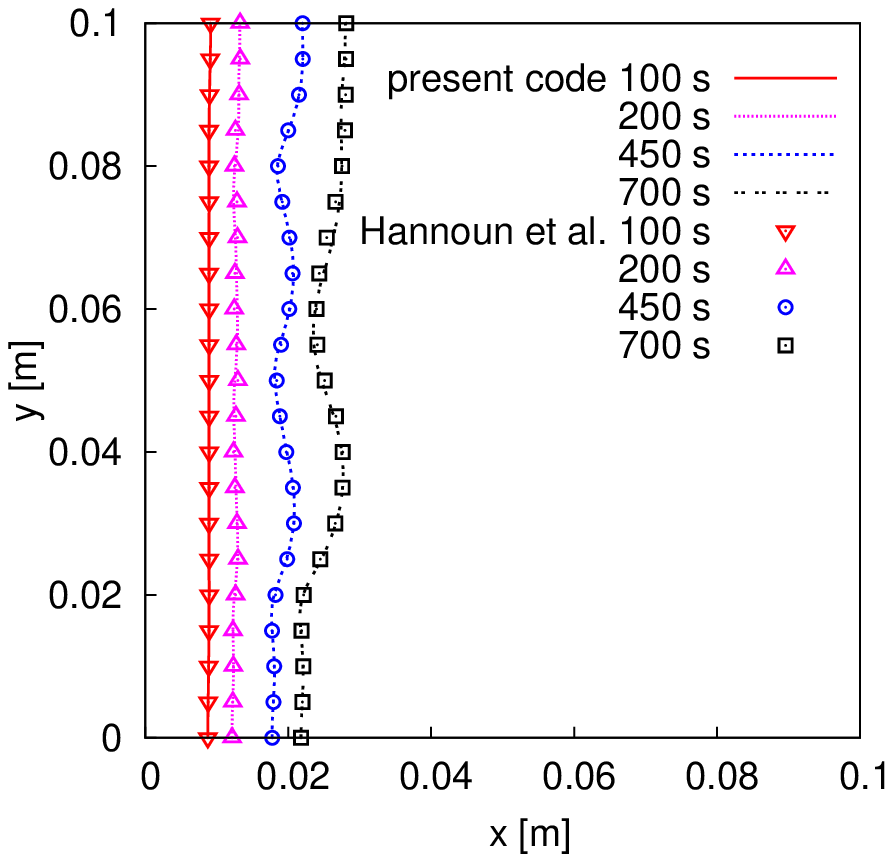} 
 \includegraphics[width=0.49\textwidth]{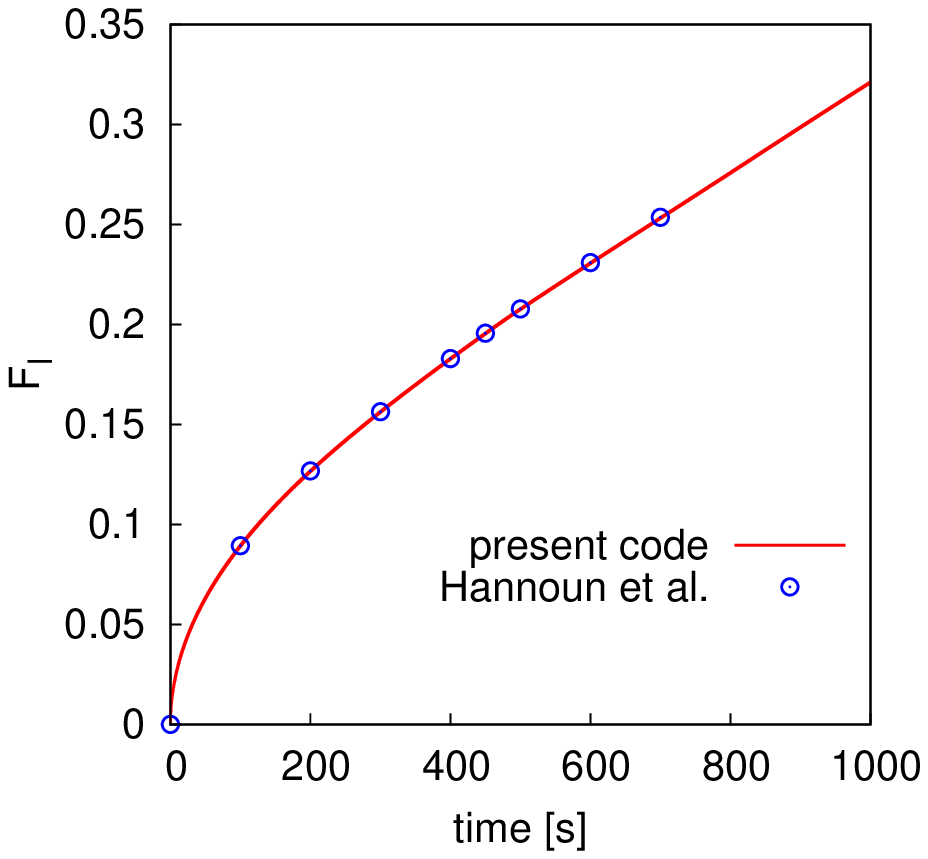}
\caption{\it \small  Comparison of the obtained melting front positions at several instants  (left) and of the evolution of the total liquid fraction in the enclosure (right) with the benchmark results of Hannoun et al. \cite{hannoun2005}.} \label{fig:bench}
\end{center}
\end{figure}

We carry out a numerical simulation in the same conditions as those described in \cite{hannoun2005} with the same material properties using a uniform grid 200 $\times$ 200 in size. In Fig. \ref{fig:bench}, we plot the melting front at different times. These lines correspond to the isolines of $f=0.5$. The melting fronts at times $450\ s$ and $700\ s$ present a wave-like shape that is due to the presence of three and two fluid recirculations (or rolls) respectively. The top two rolls have merged into a single one at about $t=480\ s$. We also present the total liquid fraction $F_l$ in the entire square enclosure, which is calculated as follows:
\begin{equation}
  F_{l} = \dfrac{1}{\sum_{c} V_{c}}\left(\sum_{c} V_{c} f_{c}\right), 
\label{eq:monitor1}
\end{equation} 
where $V_c$ is the volume of a computational cell $c$, and the summation is over all cells in the computational domain.

All these results are compared to those in \cite{hannoun2005} and exhibit satisfactory correspondence.

\subsection{Mesh size-dependence study}\label{sec:mesh}

To choose the mesh size with the best compromise between accuracy and computational cost, we conduct a mesh size-dependence study to determine the number of computational cells of the mesh needed to achieve satisfactory accuracy. The rectangular enclosure (\textbf{b}) undergoing an unsteady melting process in the same conditions as those described in section \ref{sec:FL} is chosen as an example to explain the mesh selection process. Three meshes ($m_1$, $m_2$ and $m_3$) of increasing size that contain 6000, 11660 and 24000 cells, respectively, are considered. The results obtained with the three meshes are compared based on the time evolution of the mean values and on the instantaneous spatial fields (i.e., melting front position, temperature and liquid fraction).

\begin{figure}[tbp]
  \begin{center}
\includegraphics[scale=0.9]{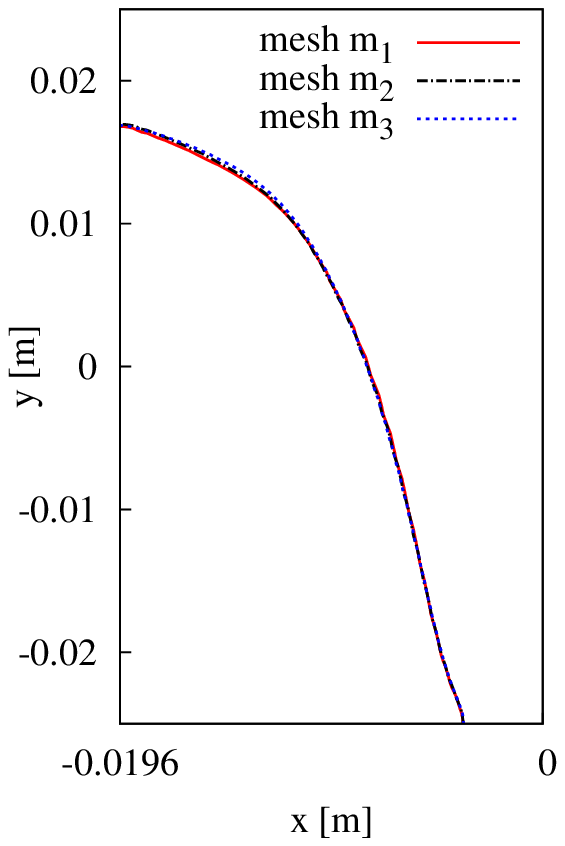} 
\includegraphics[scale=0.9]{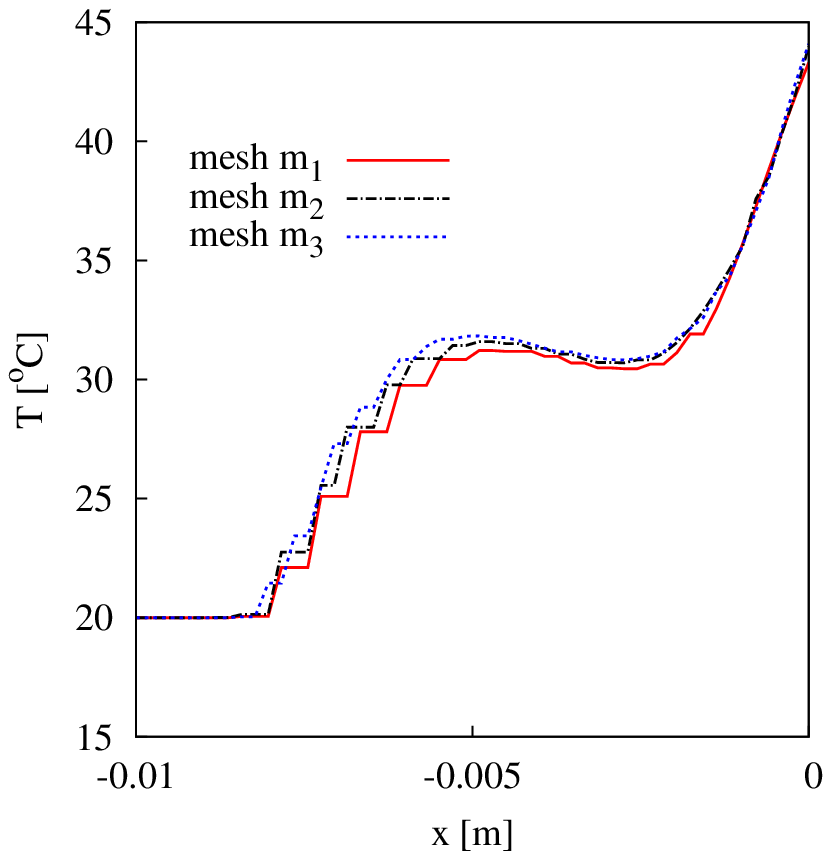} \\
(a)\hspace{6cm} (b)
\caption{\it \small Comparison of results obtained by three meshes at $t=200 \  s$: (a) melting front positions and (b) temperature along the $y=0$ cutline in the liquid phase.}\label{fig:meshstudy}
  \end{center}
\end{figure}

Fig.  \ref{fig:meshstudy}(a) gives the form of the melting front obtained by the three meshes at time $t=200\ s$. These fronts are obtained by plotting the $f=0.5$ isovalue line of the liquid fraction field. The fronts obtained by the three meshes are almost superimposed on each other, which indicates that the mesh size has little influence on the front position. More significant differences are observed in the temperature field, as seen in Fig. \ref{fig:meshstudy}(b), which shows the variation of the temperature along a horizontal axis passing through the center of the enclosure. For clarity, the plot is limited to the liquid region, where the temperature is not constant. However, the results of the three meshes are comparable. To achieve a quantitative comparison, we consider 50 points along this cutline and calculate the local error in the temperature at each point $j$ as follows:  $e_j^i = (T^{i}_j - T^{3}_j)/T^{3}_j \times 100$, where $i=1, 2$ stands for meshes $m_1$ and $m_2$; the results are compared to those of the finest mesh $m_3$. Thus, the mean value of the error is $ \bar e\ ^i = \sum_j e_j^i/50$, and the maximum error is $e^i_{max}=\max_j e_j^i$. They are both reported in Tab. \ref{tab:local_err}.

\begin{table}[th]
\caption{Relative error of the local values of $T$ along the central horizontal cutline at $t=200\ s$ using the results given by the finest mesh $m_3$ as the baseline. \label{tab:local_err}}
\centering
\begin{tabular}{c c c}
\hline
   mesh            &  $ \bar e$ & $e_{max}$ \\
\hline
 $m_1$ 		&  2.2\%    & 8.1\%    \\
$m_2$  		&   1.2\%    & 6.1\% \\
\hline
 \end{tabular}  
\end{table}

In addition to the spatial results, we compare the time evolution of the total liquid fraction in the enclosure $F_l$ (Eq. (\ref{eq:monitor1})) and the mean temperature $T_{sm}$ of the surface $S_h$, which is calculated as follows:
\begin{equation}
  T_{sm} = \dfrac{1}{\sum_{f\in S_h} A_{f}}\left(\sum_{f\in S_h} A_{f} T_{f}\right) 
\label{eq:Tsm},
\end{equation} 
 where $A_f$ is the area of a mesh face $f$ located on the surface $S_h$. Relative errors of meshes $m_1$ and $m_2$ compared to $m_3$ are calculated at each time step for $F_l$ and $T_{sm}$ over $1000\ s$. The time average and maximum values of these errors are reported in Tab. \ref{tab:mean_err} 
   
 \begin{table}[th]
 \caption{Mean and maximum values of the relative error of $F_l$ and $T_{sm}$ for a duration of $1000\ s$. \label{tab:mean_err}.}
 \centering
\begin{tabular}{c c c c c}
\hline
   mesh            &  $ \bar e(F_l)$ & $e_{max}(F_l)$&$ \bar e(T_{sm})$&  $e_{max}(T_{sm})$\\
\hline
$m_1$ 		&  0.11\%    		& 0.75\%    		&0.68\% 			& 4.34\%\\
$m_2$ 		&  0.06\%    		& 0.40\% 		& 0.09\% 			& 1.43\%\\
\hline
 \end{tabular}  
\end{table}

It is evident from Tabs. \ref{tab:local_err} and \ref{tab:mean_err} that the results given by mesh $m_2$ are sufficiently close to those given by mesh $m_3$, which is two times finer. Thus, the size of mesh $m_2$ was adopted to perform the computations presented in this work, and all the meshes generated for the five enclosures have approximately 11,000 cells.

\section{Results}

The computations for the five enclosures were conducted assuming the same conditions and using identical numerical parameters. The time step size was fixed to a relatively small value $\Delta t=10^{-2}s$, and at each time step, the convergence criteria of the SIMPLE algorithm were fixed at $10^{-4}$ for the velocity residuals and $10^{-6}$ for the temperature residual. The latter criterion of convergence was fixed to a lower value in order to achieve a higher level of energy conservation in its both sensible and latent forms as they are of different orders of magnitude, furthermore, the released energy is tightly related to the amount of melted PCM. The residuals are calculated from Eq. (\ref{eq:algeb}) as follows:
\begin{equation}
  Res_\phi = \sum_c\left(  a_c \phi_c +\sum_{N_b} a_{N_b} \phi_{N_b} - b_{c}\right) \diagup  \sum_c a_c \phi_c,
\end{equation}
where $n_c$ is the number of cells in the computational domain. Within each SIMPLE iteration, the liquid fraction update algorithm is stopped when the difference between two successive $f$ values is less than $10^{-4}$. At the first SIMPLE iteration of a time step, a few iterations are needed for $f$ to converge (less than 3), and in the remaining SIMPLE iterations, only one iteration is needed.  For numerical stability reasons, the imposed heat flux rate at the surface $S_h$ is progressively raised during the first 5 seconds to reach its constant value $q'' = 10^4\ W.m^{-2}$. This practice enables easier convergence for the early time steps.

\subsection{Velocity field and melting front evolution}

We show the flow patterns represented by velocity vectors in the five studied enclosures in Fig. \ref{fig:vect}. This figure also gives the position of the melting front (or the solid phase zones) at five different instants during the process. At $t=50\ s$, a thin vertical layer of PCM has melted, and it has roughly the same size and the same form in all the enclosures. The flow field has a regular form with ascendant and descendant parallel fluid currents. Up to this time, the five enclosures have the same behavior, and the geometry does not play a role. At $t=100\ s$, more PCM has melted, especially near the upper part of the surface $S_h$. The behavior of the enclosures then differs visibly: the enclosures (\textbf{a}) and (\textbf{b}) that do not exceed the surface $S_h$ have a melting front that extends more horizontally due to the presence of the upper boundary. Thus, the shapes of melted zones are comparable between enclosures (\textbf{a}) and (\textbf{b}) and among (\textbf{c}), (\textbf{d}) and (\textbf{e}).

At $t=200\ s$, the melted PCM zone in the enclosure (\textbf{e}) extends upward with a round shape, and as a consequence, more solid PCM is exposed to the flow current of the liquid for this enclosure than the others (the liquid-solid interface is the most extended for enclosure (\textbf{e})). From this instant on, the melting fronts in the five enclosures present an inclined curve because the melting is more advanced in the upper regions due to the thermal stratification (see Fig. \ref{fig:Tiso}). 
The flow patterns in the enclosure (\textbf{e}) show the formation of recirculation above the surface $S_h$. This recirculation, visible at $t=200\ s$ (Fig. \ref{fig:vect}), is in fact an unsteady oscillatory phenomenon with a period less than 2 s, which is why it is not visible in the figure when $t=300\ s$. These oscillations are induced by the formation of a large liquid space above the surface $S_h$. At the upper extremity of $S_h$, the hot flow stream rising along the surface oscillates between two orientations: a vertical orientation, as observed at $t=300\ s$, and a roughly horizontal orientation, as observed at $t=200\ s$ or $t=500\ s$, which results in a secondary recirculation in the upper-right corner. This unsteady behavior starts as early as $t\approx 150\ s$ for enclosure (\textbf{e}) but much later ($t \approx 800\ s$) for enclosure (\textbf{d}) because the upper liquid space is narrower. The flows in the three other enclosures do not show any oscillatory characteristics. At $t=500\ s$, the melting in all the enclosures is almost complete. We can visually observe that enclosures (\textbf{a}) and (\textbf{e}) contain the least amount of solid PCM. This observation is confirmed quantitatively in Fig. \ref{fig:fl}.

To give more quantitative details about the velocity field in the enclosures, we plot the vertical component of the velocity vector along the horizontal axis passing through the center of the surface $S_h$ in Fig. \ref{fig:Vm_x}. Initially (up to $t=100\ s$), the velocities are similar, so they are not shown. At $t=200\ s$, we observe substantial differences in the maximum value of the velocity near the boundary $S_h$, where enclosure (\textbf{e}) exhibits the highest value ($6.5\times 10^{-3}\  m.s^{-1}$), and enclosure (\textbf{a}) exhibits the lowest value ($4.7\times 10^{-3}\ m.s^{-1}$). The $x$ positions of the positive maximum are nearly the same, whereas in the fluid zone near the melting front, the downward negative velocities have different profiles due to the differences in the position of the front.
As the process continues ($t=300\ s$ and $t=500\ s$), the velocity level decreases progressively, and a large zone with almost zero vertical velocity appears in the center of the enclosures. The maximum value in enclosure (\textbf{e}) at $t=500\ s$ is $3.1\times 10^{-3}\ m.s^{-1}$. This velocity decrease is due to the reduction of the quantity of solid PCM, which is at the low temperature $T=T_m$, in the enclosures compared to the initial amounts. As a result, a weaker temperature gradient exists in the liquid which results in weaker buoyancy forces.

\begin{figure}[p]
  \begin{center}
\includegraphics[scale=0.18]{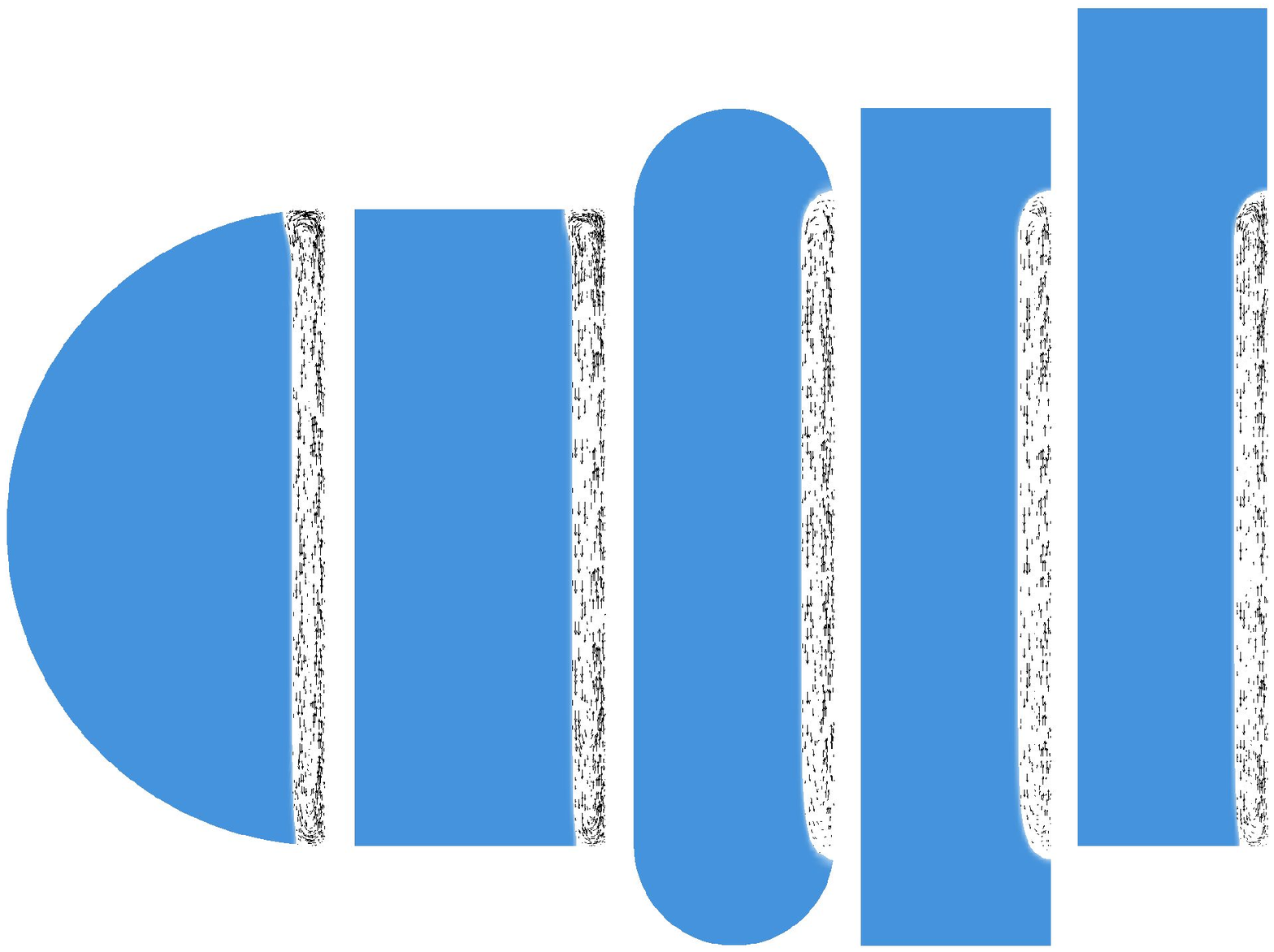} \\
\includegraphics[scale=0.18]{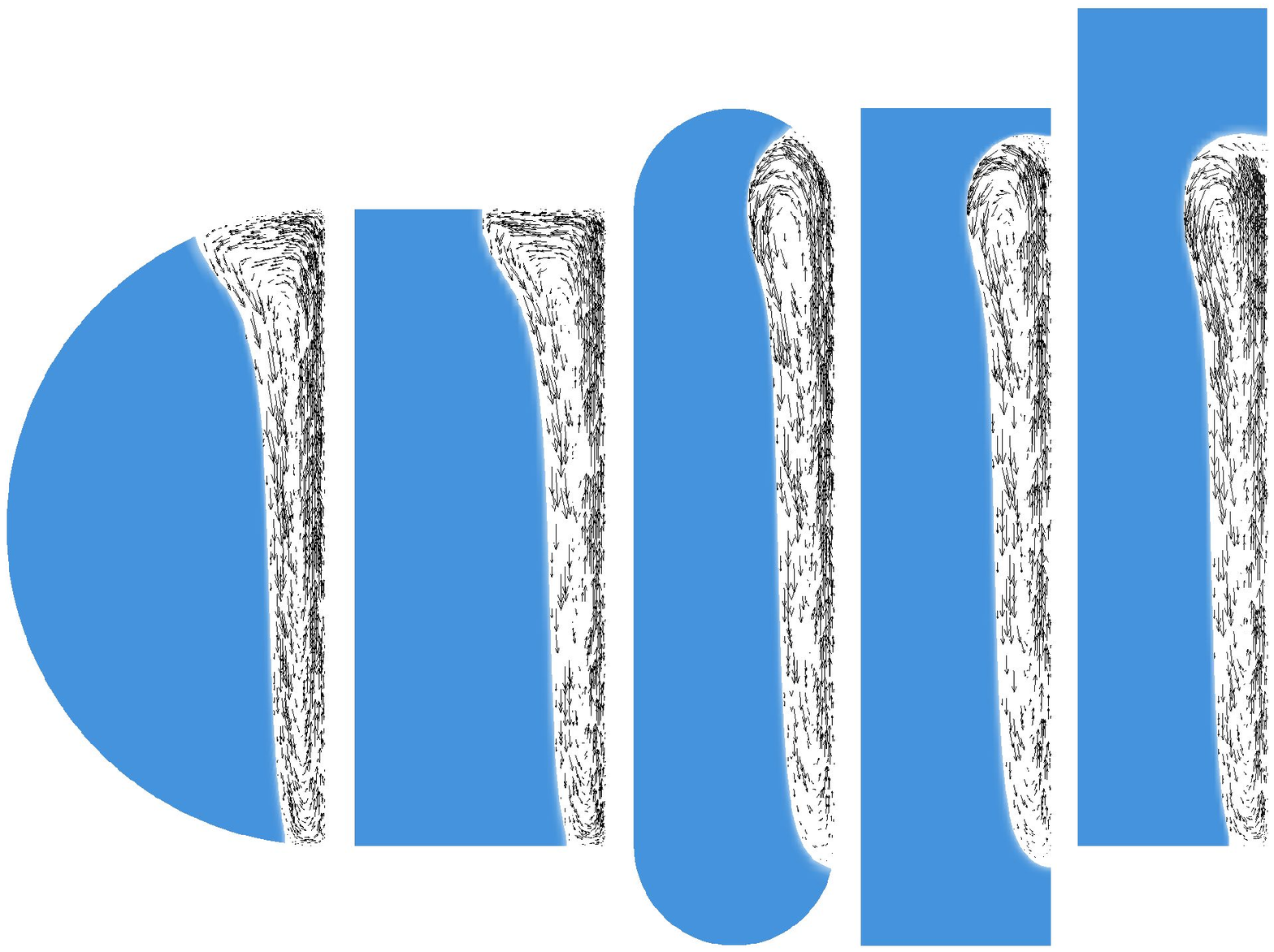} 
\caption{\it \small  Velocity vectors and melting fronts for the five enclosures after 50 and 100 s. }\label{fig:vect}
  \end{center}
\end{figure}
\addtocounter{figure}{-1}
\begin{figure}[p]
  \begin{center}
\includegraphics[scale=0.18]{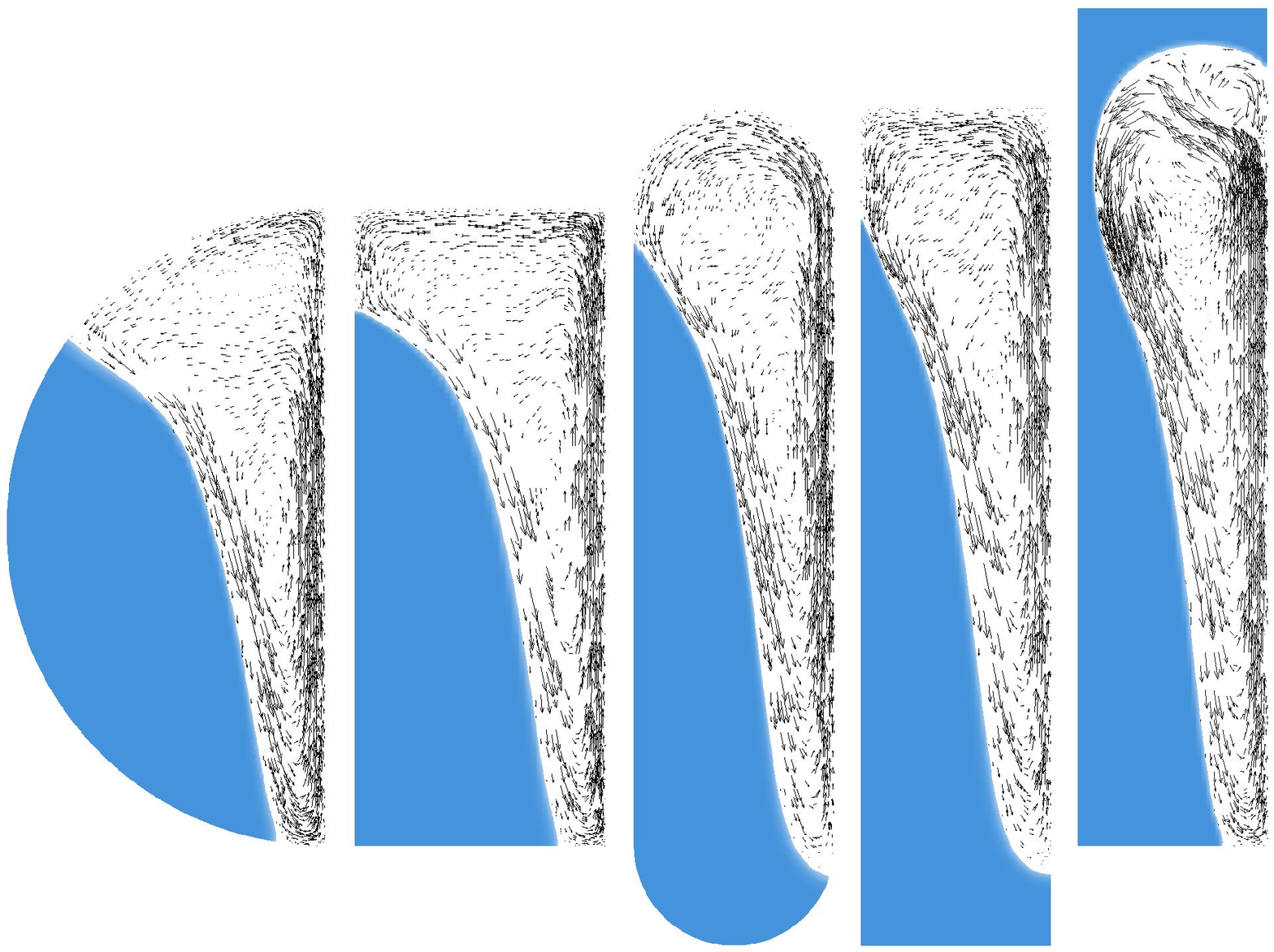} \\
\includegraphics[scale=0.18]{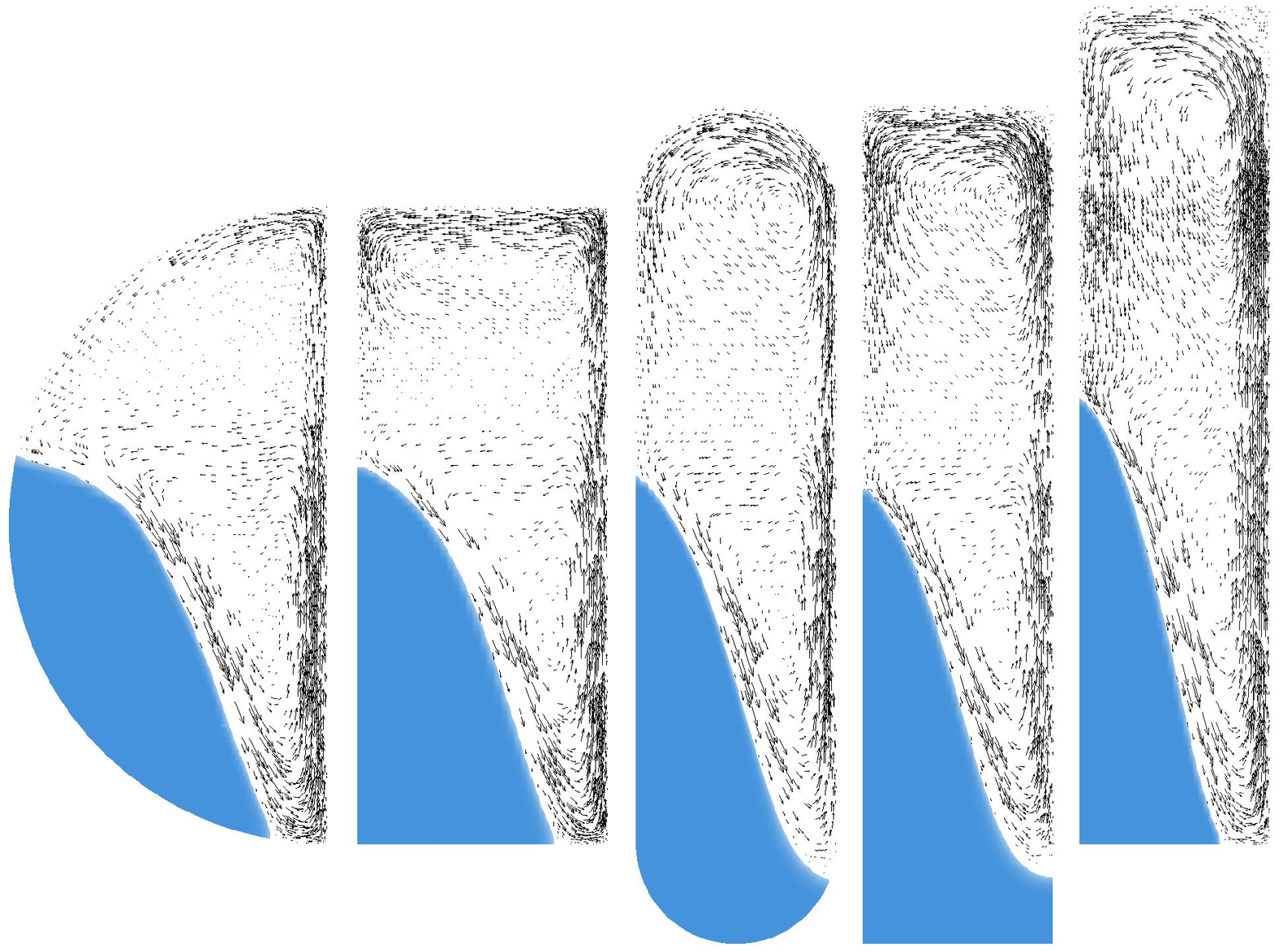} 
\caption{\it \small (Cont.) Velocity vectors and melting fronts for the five enclosures after 200 and 300 s. }%\label{fig:vect}
  \end{center}
\end{figure}
\addtocounter{figure}{-1}
\begin{figure}[tp]
  \begin{center}
\includegraphics[scale=0.18]{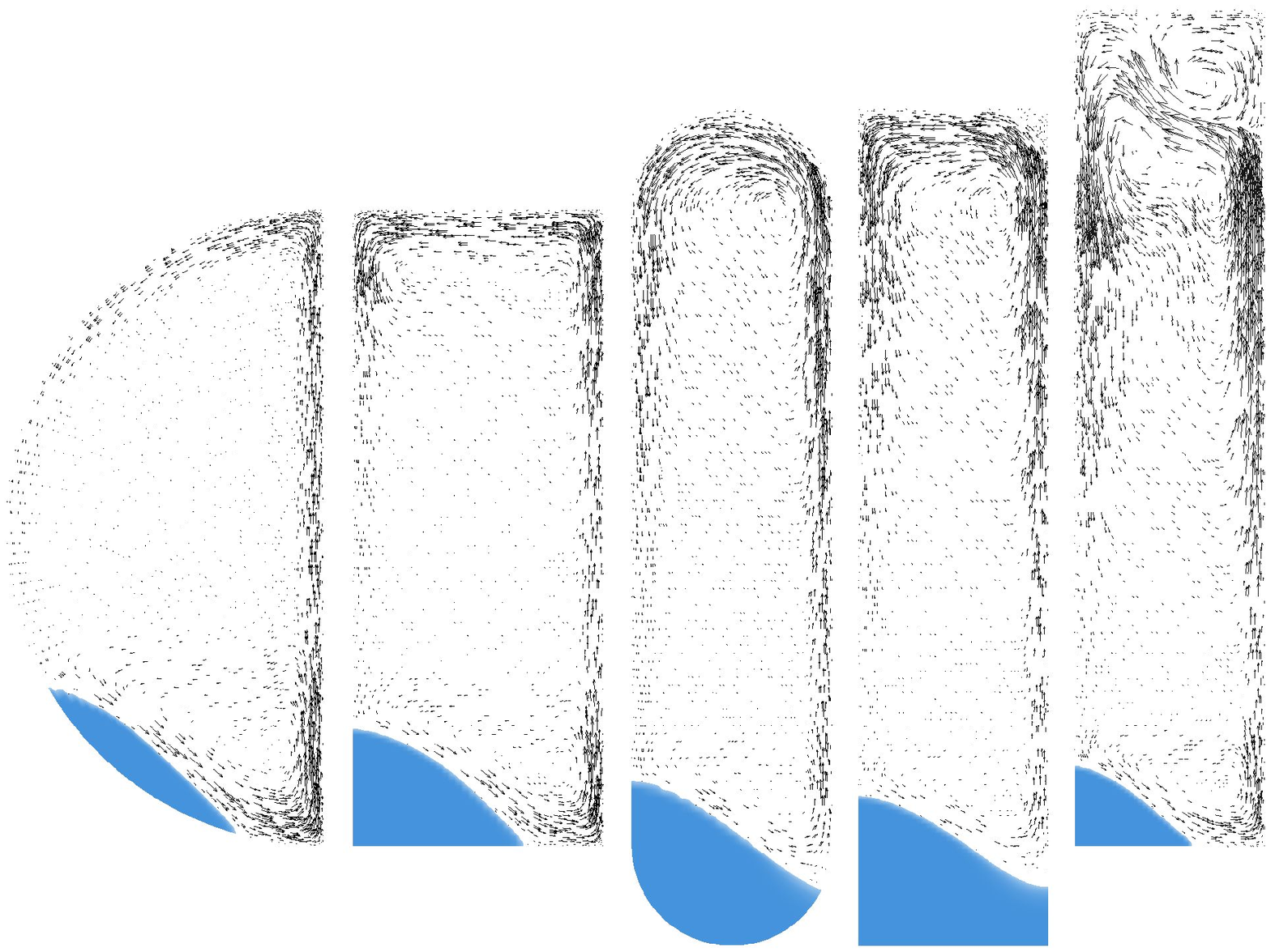} 
\caption{\it \small  (Cont.) Velocity vectors and melting fronts for the five enclosures after 500 s. }%\label{fig:vect}
  \end{center}
\end{figure}
%\FloatBarrier

\begin{figure}[p]
  \begin{center}
\includegraphics[scale=0.5]{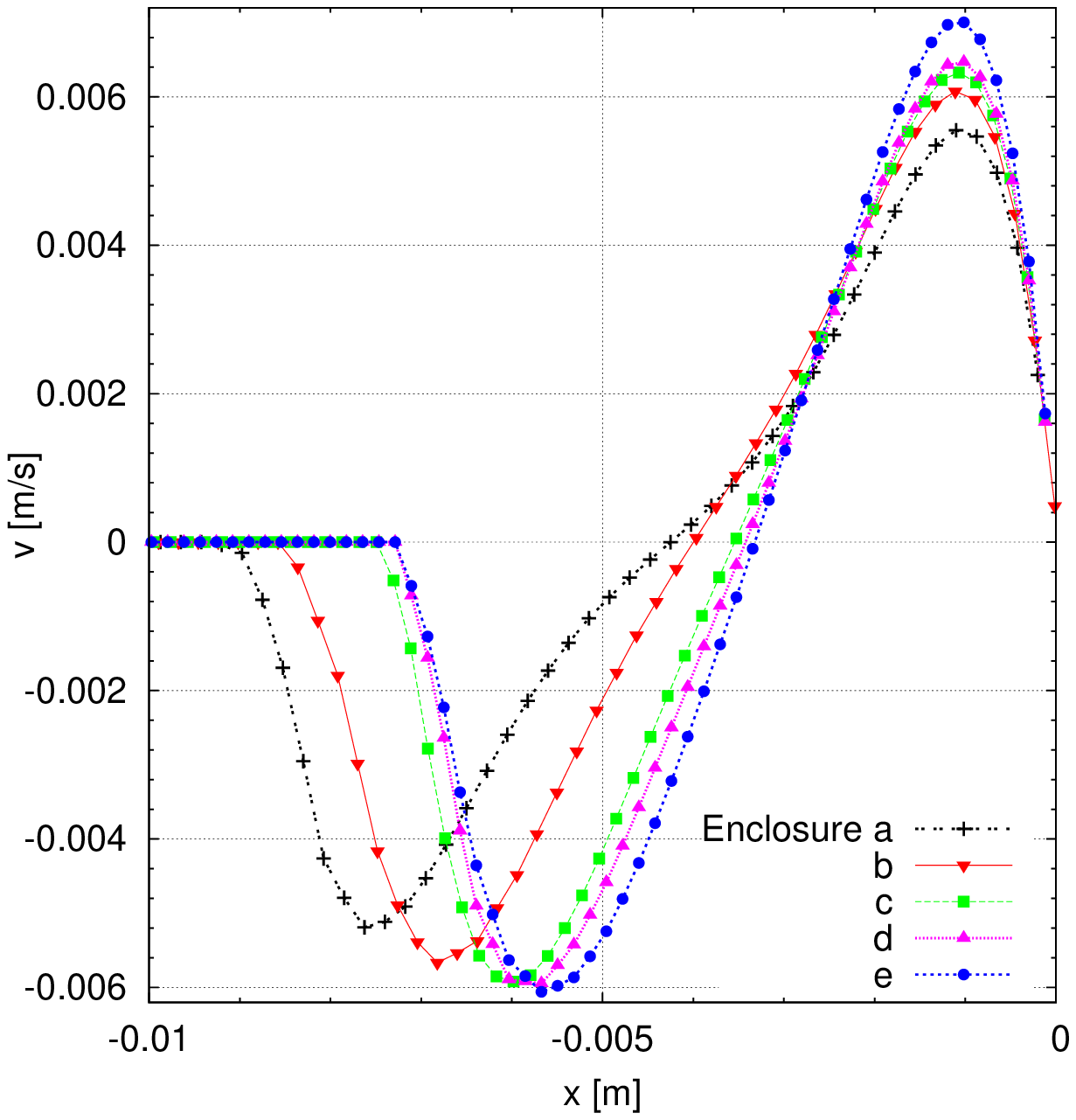} \hspace{5mm}
\includegraphics[scale=0.5]{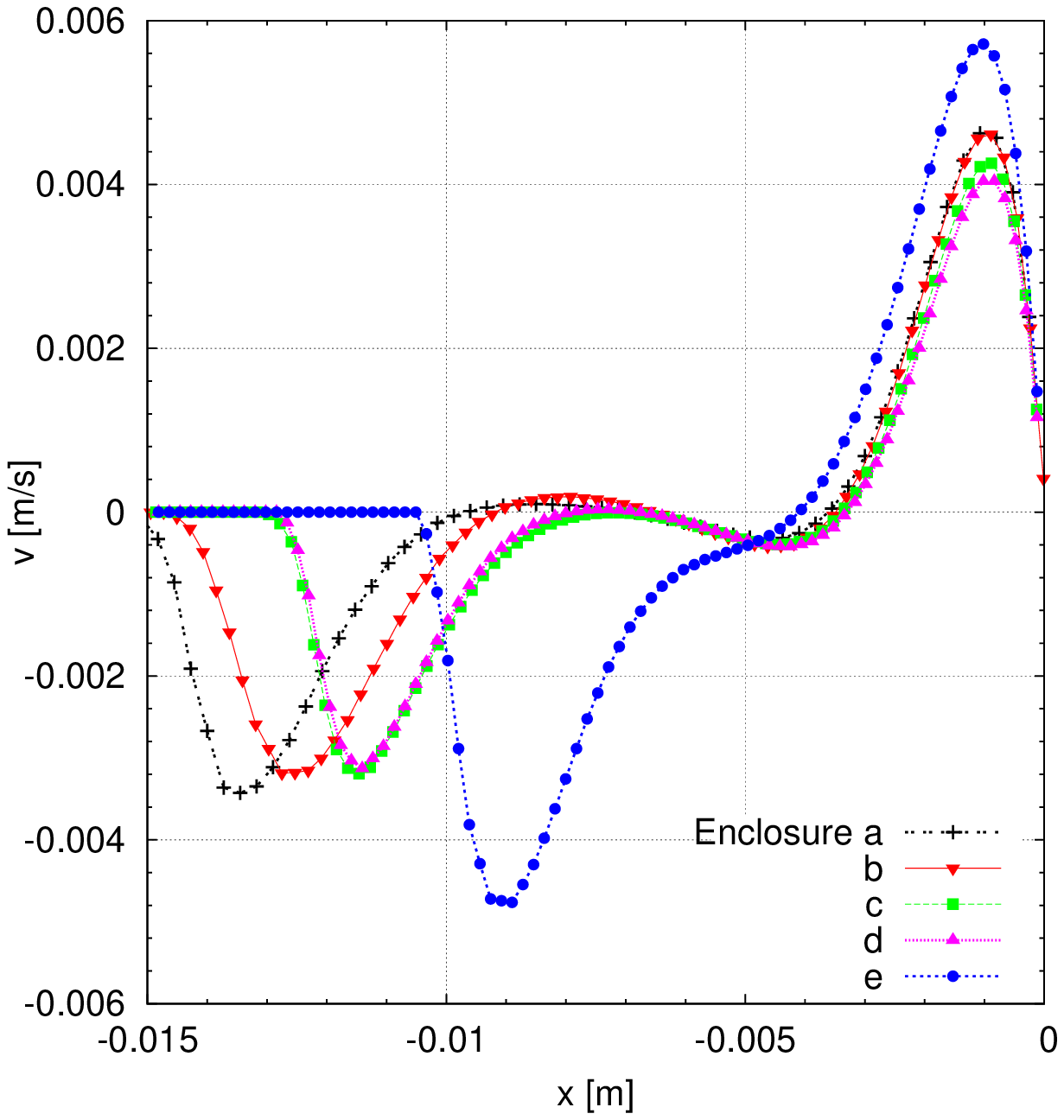} \\
(a) \hspace{7cm}(b)\\
\includegraphics[scale=0.5]{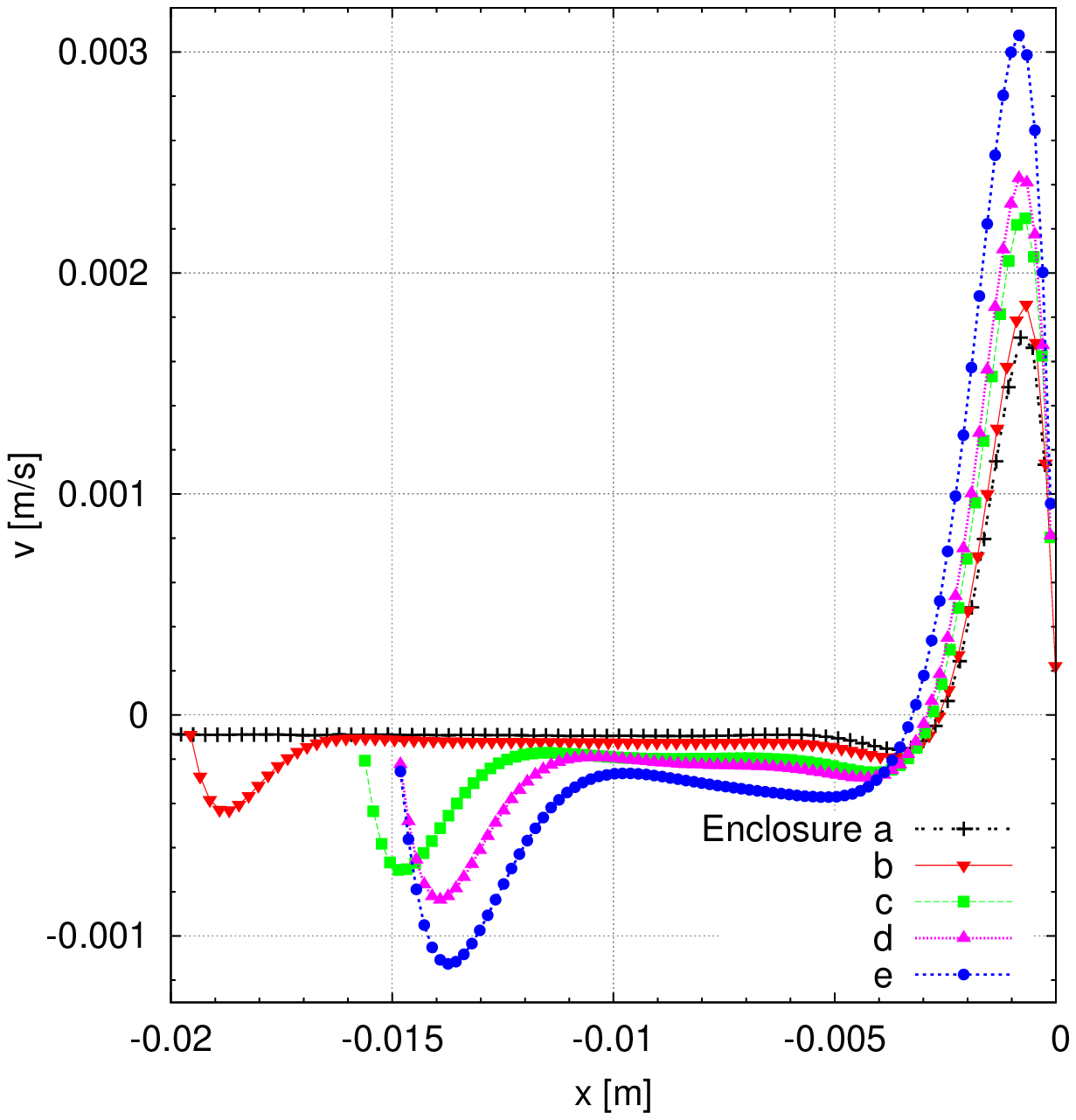} \\(c)
\caption{\it \small Vertical velocity components along the horizontal axis passing through the center of the surface $S_h$ for the five enclosures after 200 s (a), 300 s (b) and 500 s (c). }\label{fig:Vm_x}
  \end{center}
\end{figure}

\subsection{Temperature field in the enclosures}

\begin{figure}[p]
  \begin{center}
\includegraphics[width=0.49\textwidth]{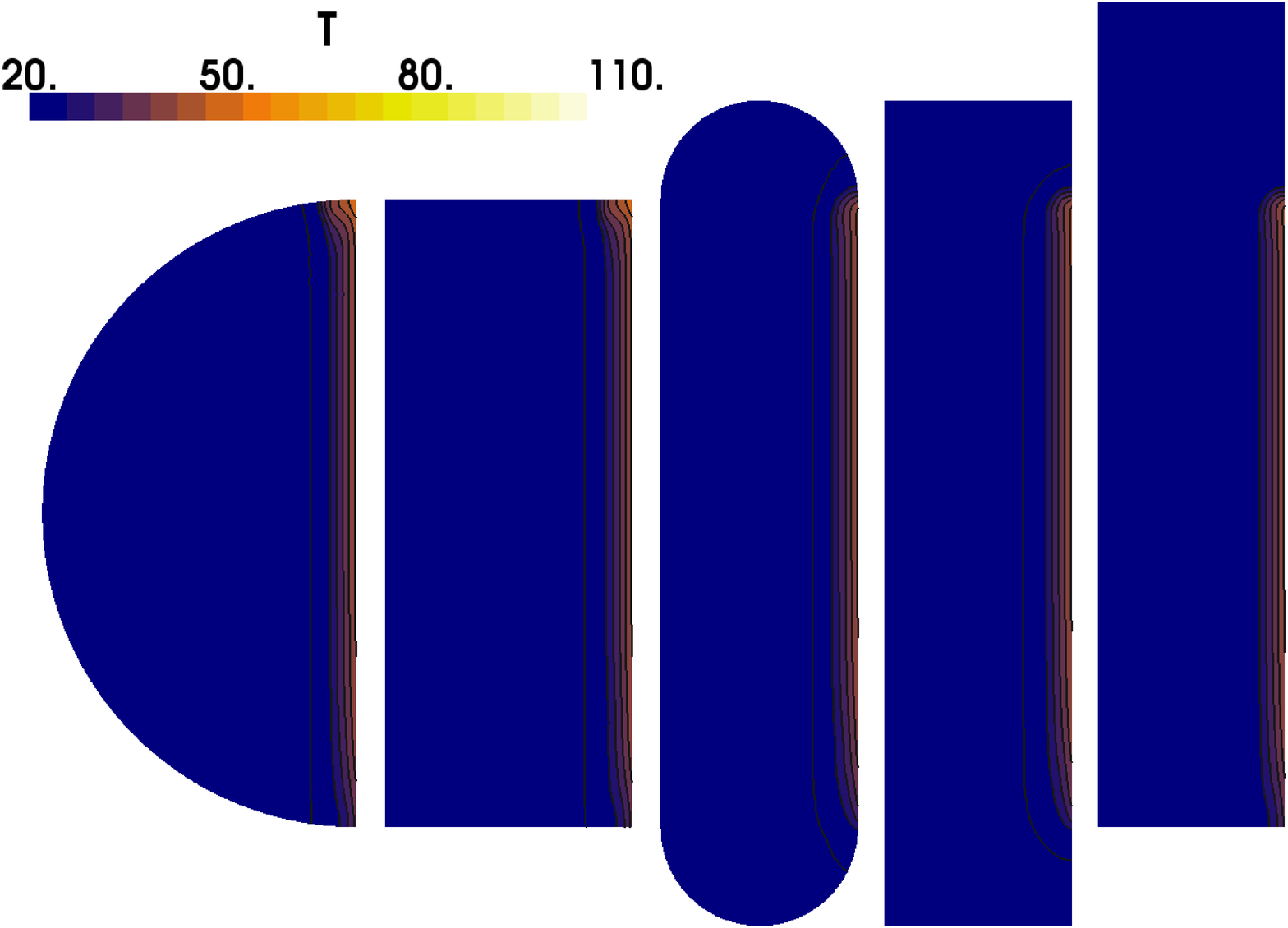} 
\includegraphics[width=0.49\textwidth]{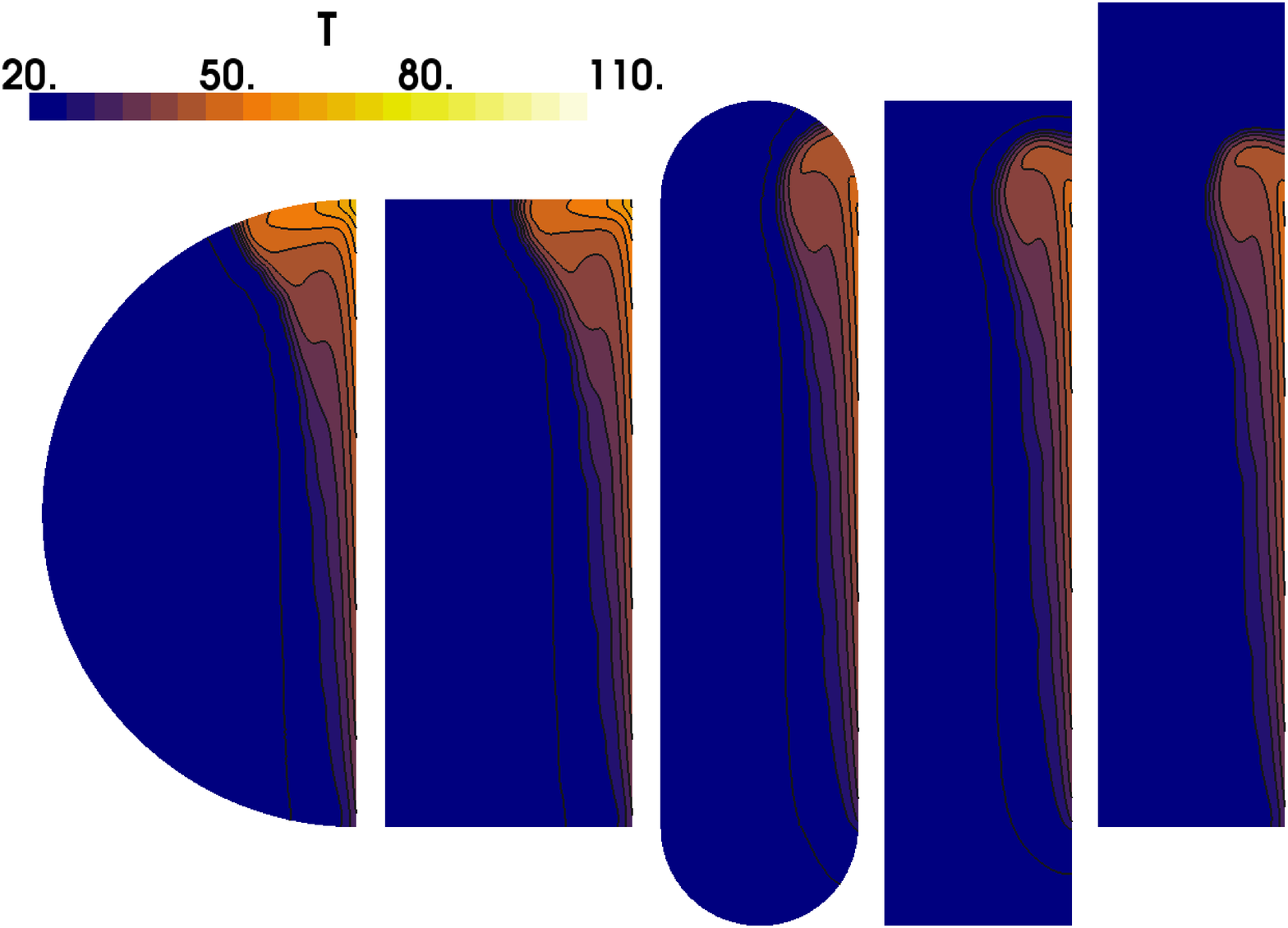} \\
\includegraphics[width=0.49\textwidth]{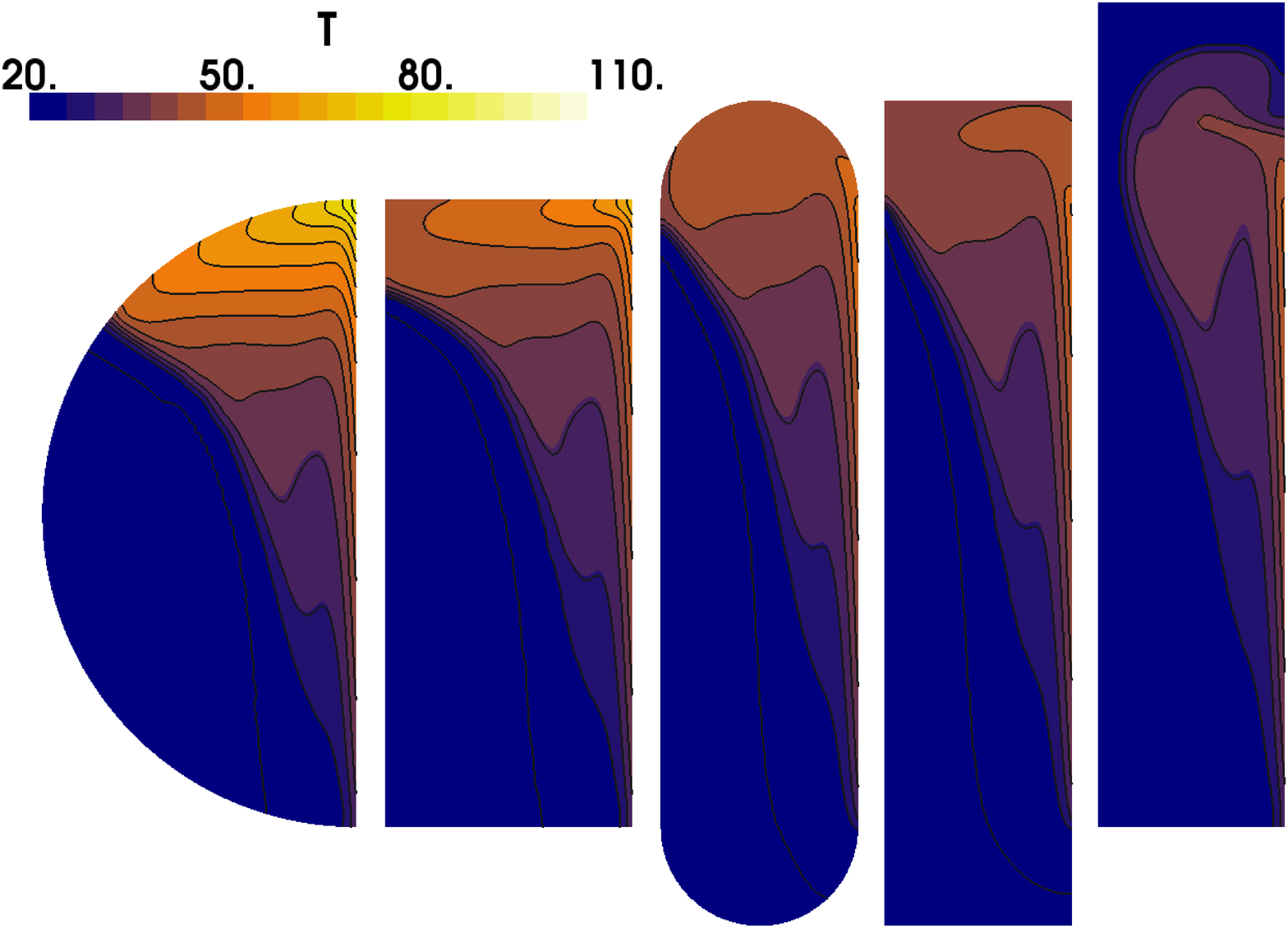} 
\includegraphics[width=0.49\textwidth]{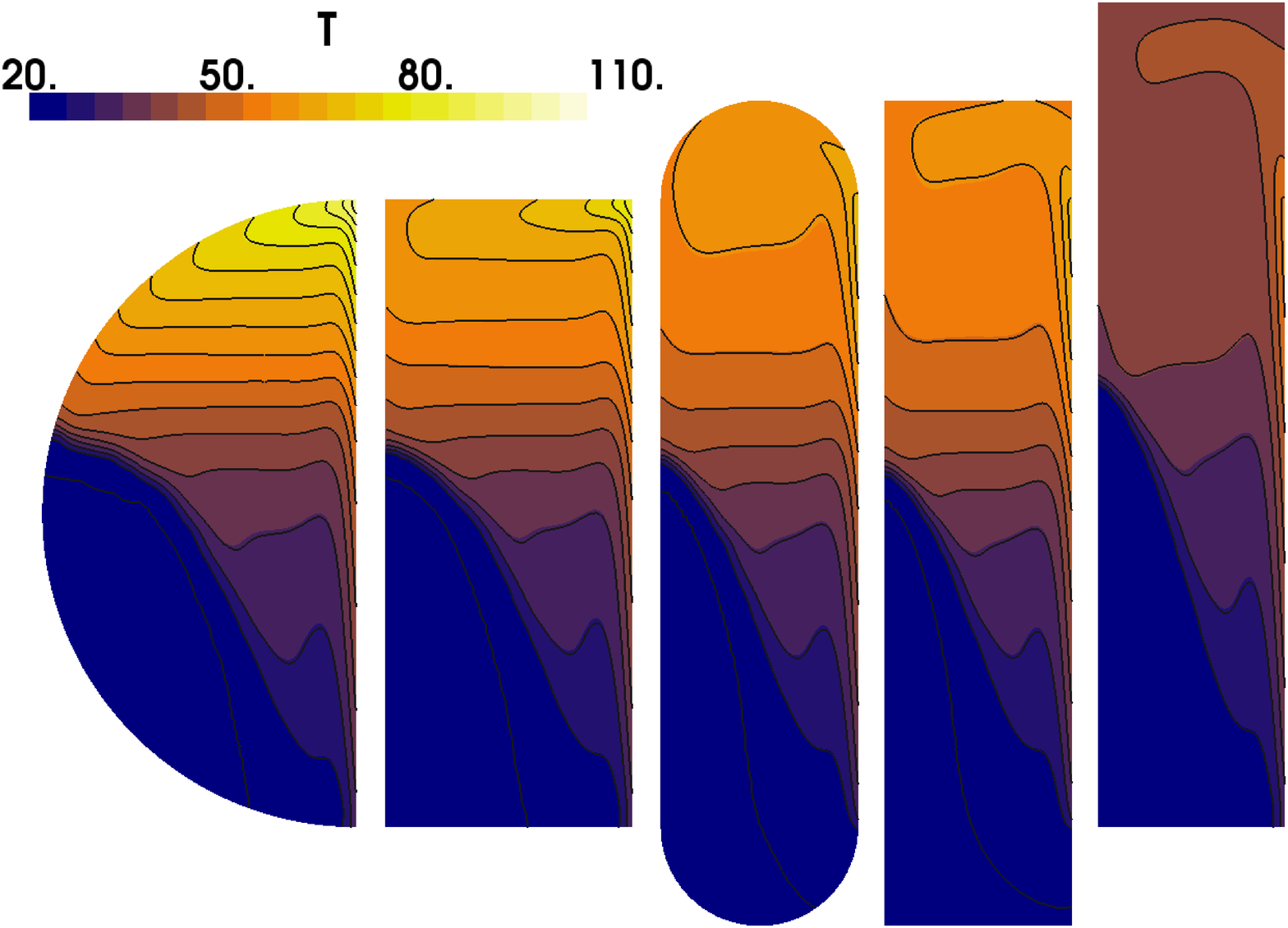} \\
\includegraphics[width=0.49\textwidth]{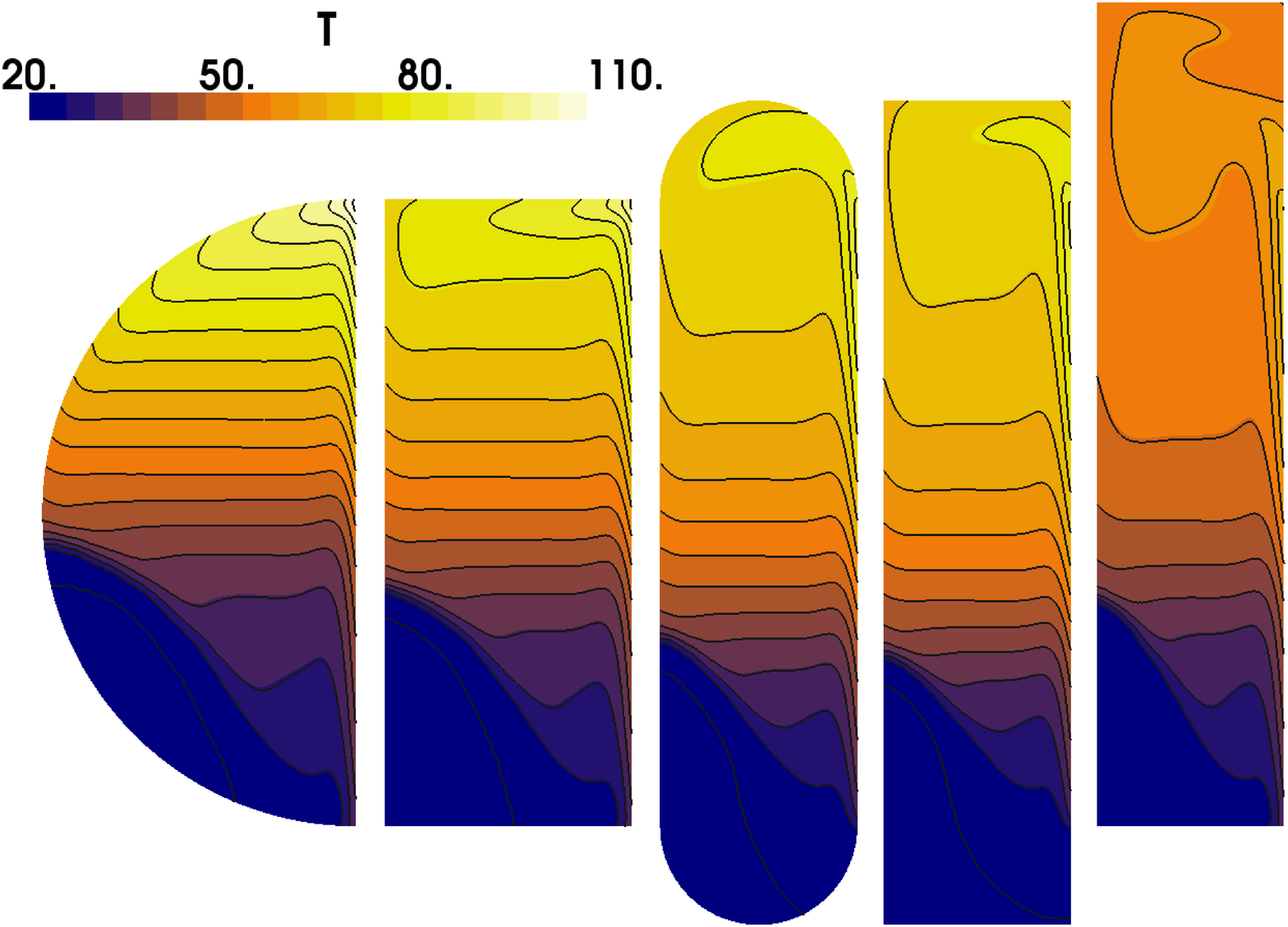} 
\includegraphics[width=0.49\textwidth]{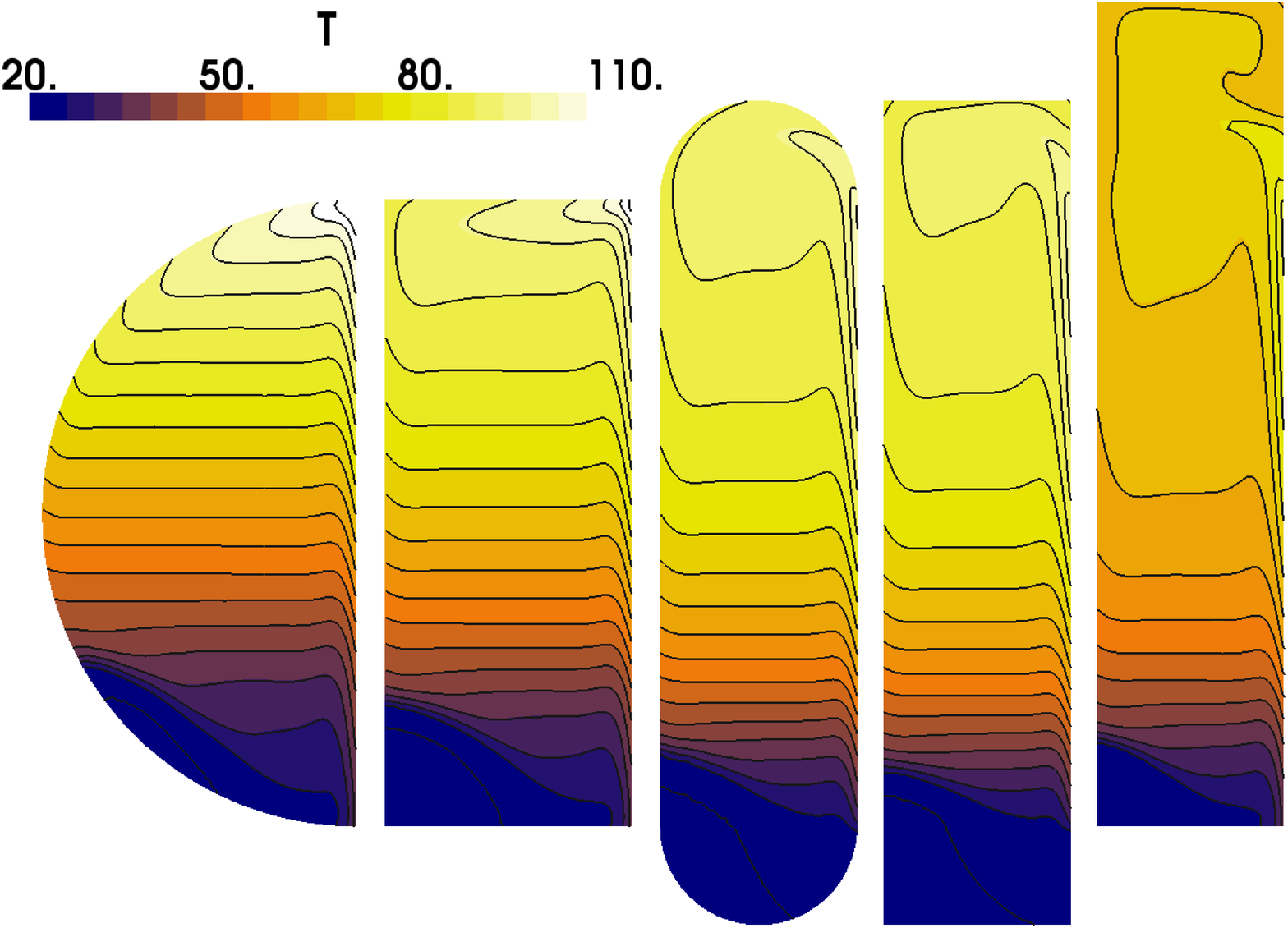} 
\caption{\it \small  Temperature fields in the five enclosures after 50, 100, 200, 300, 400 and 500 s. }\label{fig:Tiso}
  \end{center}
\end{figure}

Fig. \ref{fig:Tiso} shows the temperature field at six different times during the melting process. As early as $t=200\ s$, we observe the establishment of thermal stratification in the melt, which is slightly perturbed in the upper region by the upward fluid currents and near the lateral boundaries by the convective heat exchange with the external media. Higher temperatures were observed in the liquid of enclosure (\textbf{a}), and lower temperatures were observed in enclosure (\textbf{e}). Quantitative changes in temperature are given in the next section. As for the melting front position, we observe similarities in the temperature fields between enclosures (\textbf{a}) and (\textbf{b}) and between enclosures (\textbf{c}) and (\textbf{d}), whereas enclosure (\textbf{e}) exhibits a distinct temperature pattern. 

\subsection{Evolution of the parietal temperature}

\begin{figure}[btp]
  \begin{center}
\includegraphics[width=0.7\linewidth]{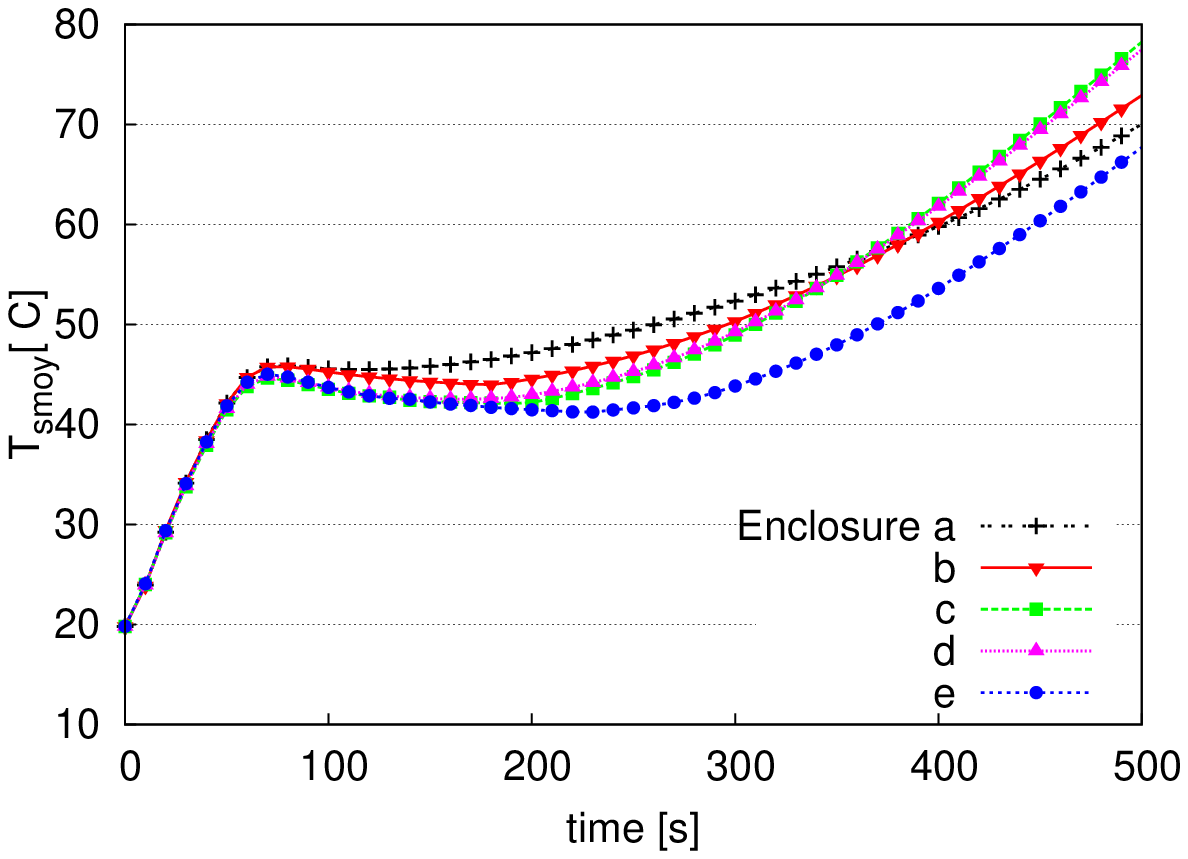}
\caption{\it \small  Evolution of the mean temperature of the surface $S_h$. \label{fig:Tsmoy}}
\end{center}
\end{figure}

The evolution of $T_{smoy}$, the mean temperature of the surface $S_h$ (Eq. (\ref{eq:Tsm})), gives an overview of the impact of the flow and melting kinetics on the global temperature level (Fig. \ref{fig:Tsmoy}). During the first minute, the mean temperature evolves identically and linearly for all the enclosures. This first phase is governed essentially by heat conduction; thus, it gives the same results for all five enclosures. After this phase of steep augmentation, $T_{smoy}$ continues to increase in enclosure (\textbf{a}) but with a weaker slope, but it decreases in all the other enclosures. The strongest decrease is observed for enclosure (\textbf{e}), and it lasts for approximately 250 s before increasing again. For enclosures (\textbf{b}), (\textbf{c}) and (\textbf{d}), this period is approximately 170 s. 

\begin{figure}[btp]
  \begin{center}
\includegraphics[width=0.7\linewidth]{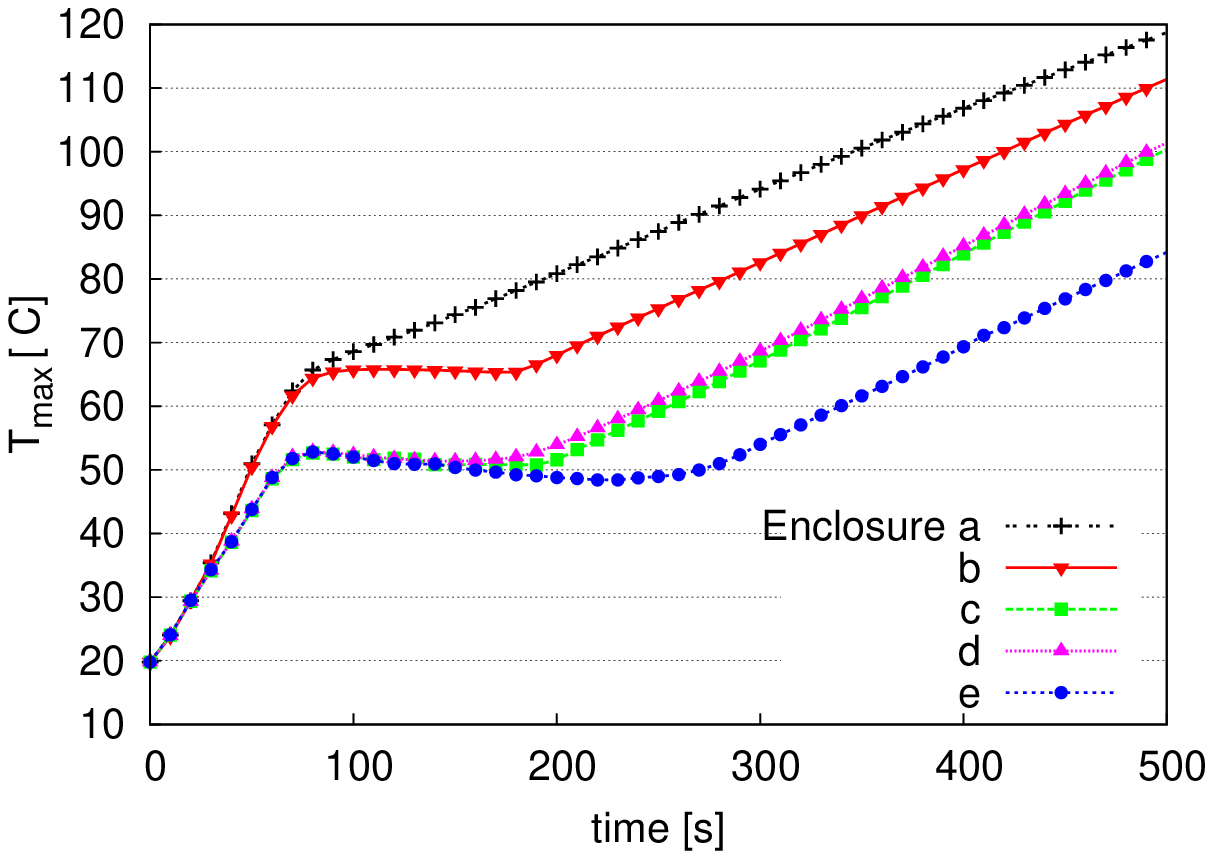}
\caption{\it \small  Evolution of the maximum temperature of the surface $S_h$. \label{fig:Tsmax}}
\end{center}
\end{figure}

Another important temperature to monitor is the maximum temperature $T_{max}$ of the surface $S_h$ because this temperature should be lower than the damage temperature of the device to be cooled. Fig. \ref{fig:Tsmax} plots the maximum temperature for the five enclosures. This maximum is always located at the top of $S_h$, as in Fig. \ref{fig:Tiso}. Its evolution shows differences between the enclosures as early as $t=30\ s$. For enclosure (\textbf{a}), $T_{max}$ exhibits two linear phases with positive slopes with inflexions at about $t=75\ s$ and $T_{max} = 65^\circ C$. The evolution of $T_{max}$ for enclosure (\textbf{b}) has a horizontal plateau at $65^\circ C$, whereas for the other three enclosures, this plateau is $15^\circ C$ lower and is slightly decreasing. The plateau for enclosure (\textbf{e}) ($\sim 200\ s$ long) is the longest. 

The plateau in the $T_{smoy}$ and $T_{max}$ curves emerges because of the establishment of natural convection currents and the presence of the front of solid PCM. The solid PCM absorbs the thermal energy transported by the flow from the hot surface in the form of latent heat, and its melting releases cold liquid at $T_m$, which increases the temperature gradient. The circular shape of enclosure (\textbf{a}) induces a larger distance between the solid PCM and the surface $S_h$. This distance is less important for enclosure (\textbf{b}), and its rectangular shape induces a plateau in the temperature profile and a lower value of $T_{max}$. When the enclosures are flatter, as for (\textbf{c}) and (\textbf{d}), the results are better. The shape of the enclosure has another effect in terms of the extent of the melting front, which is more extended for the enclosures with the lowest aspect ratio. The more geometrically extended this melting front is in the enclosure, the lower and longer the $T_{max}$ plateau. By comparing the curves of enclosures (\textbf{c}) and (\textbf{d}) with (\textbf{e}), we can conclude that an efficient way to increase the extent of the front is to locate more PCM at the top of the enclosure where it is in contact with oscillatory natural convection streams. The higher temperature liquid is located in this zone because of thermal stratification, which increases the melting rate.

\subsection{Evolution of the global PCM melting}

\begin{figure}[btp]
  \begin{center}
\includegraphics[width=0.7\linewidth]{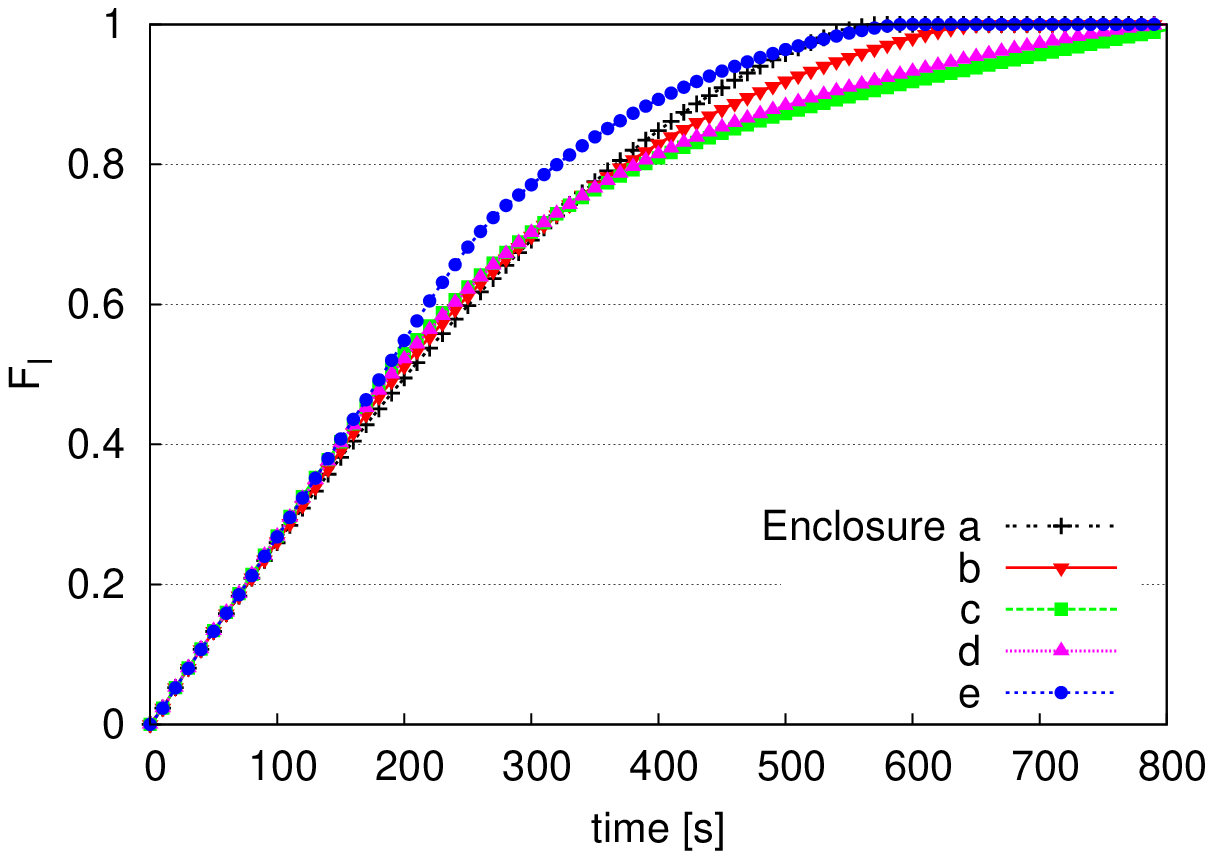}
\caption{\it \small  Evolution of the total liquid fraction in the five enclosures. \label{fig:fl}}
\end{center}
\end{figure}

The total liquid fraction $F_l$ defined by Eq. (\ref{eq:monitor1}) can be monitored to follow the evolution of  the melting in the five enclosures, as shown in Fig. \ref{fig:fl}. We observe that the curve of $F_l$  is linear and identical in all the enclosures until $t\approx 150\ s$. After this time, the rate of melting decreases in all the enclosures except for enclosure (\textbf{e}), for which the curve inflexion occurs as late as $t\approx 250\ s$. This enclosure has the fastest rate of melting, and the solid PCM completely disappears at $t\approx 550\ s$. 
Enclosure (\textbf{a}) is also completely melted at this time, and even though it shows the highest maximum surface temperature (Fig. \ref{fig:Tsmax}), its mean surface temperature is lower than those of enclosures (\textbf{b}), (\textbf{c}) and (\textbf{d}) after $t\approx 350\ s$ (Fig. \ref{fig:Tsmoy}). Fig. \ref{fig:vect} can explain these differences. The bottom left corner of the rectangular enclosure (\textbf{b}) retains the PCM solid and limits its melting. In enclosures (\textbf{c}) and (\textbf{d}), more solid PCM is trapped in the lower part of the enclosure located below the surface $S_h$. This zone is less influenced by the hot liquid streams recirculating inside the enclosure, and thus, it is more difficult to melt. As a consequence, enclosures (\textbf{c}) and (\textbf{d}) are the last ones to completely melt, approximately $250\ s$ after  enclosures (\textbf{a}) and (\textbf{e}) are completely melted.

\subsection{Parietal heat transfer}

To quantify the heat transfer at the cooled surface, we define a global Nusselt number as follows:
\begin{equation}
  Nu =\frac{q'' H}{k(T_{smoy} - T_{m})}.
\end{equation} 

\begin{figure}[btp]
  \begin{center}
\includegraphics[width=0.7\linewidth]{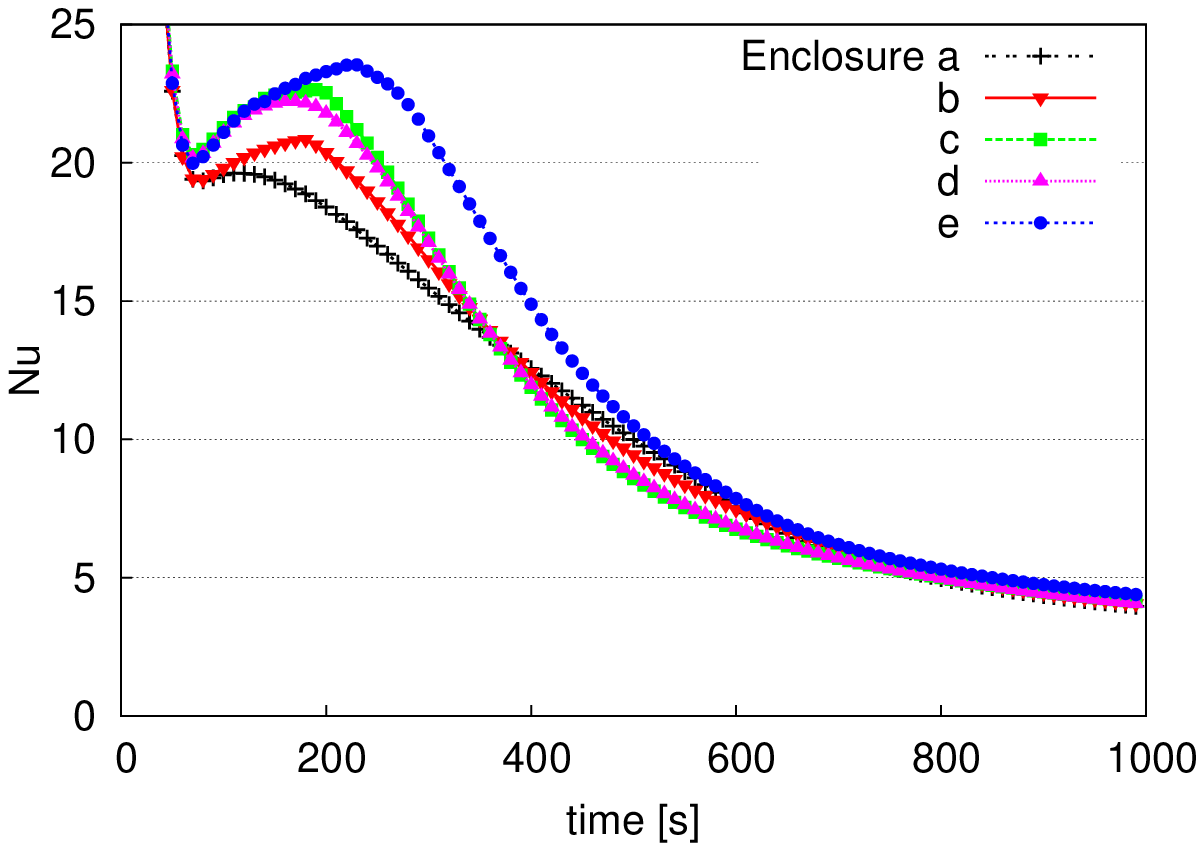}
\caption{\it \small Evolution of the parietal Nusselt number for the five enclosures. \label{fig:Nu}}
\end{center}
\end{figure}

Fig. \ref{fig:Nu} gives the evolution of the Nusselt number for the five enclosures during the melting process. After a fast transition period, where the $Nu$ values are important ($T_{smoy} - T_{m} \approx 0$), and before they drop to a value of $20$, we observe an increase toward a peak value followed by a progressive decrease toward an asymptotic value of $4$. The characteristic ``bumps'' in the curves obtained for the different enclosures have different sizes and correspond to the coupled impact of the natural convection and latent heat of the melting of the PCM on the temperature of the cooled surface $S_h$. In agreement with the results of velocity (Fig. \ref{fig:Vm_x}) and parietal temperature (Fig. \ref{fig:Tsmoy}), the most important $Nu$ value is obtained for enclosure (\textbf{e}), and the lowest is obtained for enclosure (\textbf{a}) up to $t\approx 350\ s$. After $t\approx 350\ s$,  enclosure (\textbf{a}) performs better than enclosures (\textbf{b}), (\textbf{c}) and (\textbf{d}) when the rate of fusion in these enclosures slows down. However, the most interesting behaviors expected from this mode of passive cooling are the fast reaction and limited maximum temperature. With regard to these two features, enclosure (\textbf{a}) is the worst choice and enclosure (\textbf{e}) is the best choice. We can also notice that for all the efficiency indicators ($T_{smoy}$, $T_{max}$, $F_l$ and $Nu$), the oblong enclosure (\textbf{c}) performs slightly better than the rectangular enclosure (\textbf{d}).

\section{Conclusions}

In this study, we numerically modeled the natural convection-dominated melting of a PCM inside an enclosure used to control the temperature of a surface releasing a heat flux. We investigated the impact of the shape of this enclosure and the relative position of the cooled surface on the flow and heat transfer. We developed a numerical model using a fixed grid enthalpy-porosity technique coupled with a three-dimensional flow solver based on an unstructured finite-volume method involving second-order accurate spatial and temporal numerical schemes. 

Five geometries containing the same quantity of PCM were compared. The geometries were rounded or rectangular, thick or thin, centered relative to the cooled surface or shifted upward vertically. The unsteady behaviors of the five enclosures were analyzed by examining the evolution of the liquid-solid interface and considering the forms and strengths of the natural convection flow. The fluid velocity increases for thinner enclosures that have a space above the cooled surface; thus, in the enclosure shifted upward (enclosure (\textbf{e})) the maximum velocity is achieved. In spite of the presence of this flow, strong thermal stratification exists in the melt, which causes a hot spot located at the upper extremity of the surface to appear. 

Several efficiency indicators were monitored, including the mean and maximum temperatures, the total liquid fraction and the parietal Nusselt number. All of these indicators demonstrate the importance of the effect of the geometry on cooling efficiency. If we focus on the maximum temperature indicator, we see that at as early as $t=100\ s$, the difference between the best and worst geometry choices ((\textbf{e}) and (\textbf{a}), respectively) is equal to $16.5$ °C, and this difference reaches $40$ °C when $t=300\ s$, which is of great importance for the thermal protection of the heat-dissipating devices. We can summarize the most important findings of this work as follows:
\begin{itemize}
 \item In thin enclosures, liquid PCM zones expand upward, while for wider enclosures, this expansion tends to be horizontal. Consequently, for the former enclosures, the solid cold PCM is in contact with ascending hot liquid streams, which contributes to the lowering of the global temperature.
 \item The last zones to melt in the enclosures are located at the bottom and on the side opposite to the cooled surface. Thus, PCM enclosures should be designed without corners and without zones located below this surface.
 \item A portion of the PCM should be placed above the cooled surface.
 \item The use of rounded corners has a slight positive effect on the efficiency (e.g., $Nu$ and $T_{max}$).
\end{itemize}

Based on these findings, more efficient geometries can be designed; however, in the future, the impact of the geometry choice on the inverse process, i.e., solidification,  should be taken into account and analyzed. Furthermore, the relations between the fraction of nanoparticles introduced and the thermophysical properties of the PCM such as the viscosity of the liquid phase, should be modeled and their impacts on hydrodynamics and heat transfer analyzed. These aspects constitute some of the goals of a future work.

%\bibliographystyle{elsarticle-num}
%\bibliography{../KEO_global_biblio}

\appendix

\section{Test case of natural convection between a cylinder and a square duct}\label{sec:append}

\begin{figure}[htbp]
\begin{minipage}[t]{0.55\textwidth}
 \centering
 \includegraphics[width=\textwidth]{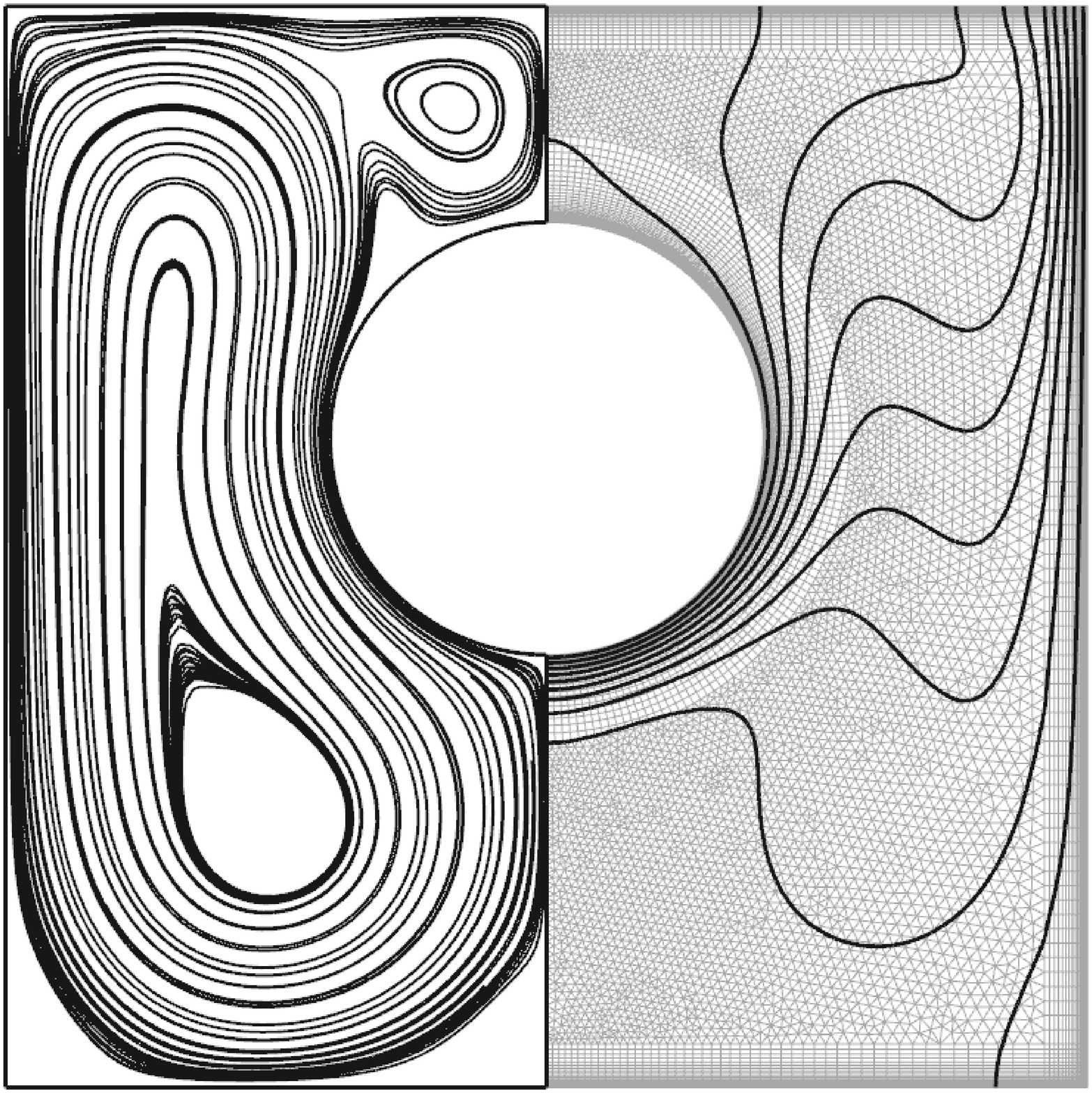}
 \caption{Calculated results of the natural convection test case of Demirdzic et al. \cite{demir1992}. Right half: the used grid and temperature contours (values from $0.05$ to $0.95$ with a step of $0.1$). Left half: streamline pattern. \label{valid_convnat}}
\end{minipage} \hfill
\begin{minipage}[t]{0.4\textwidth}
  \centering
 \includegraphics[width=\textwidth]{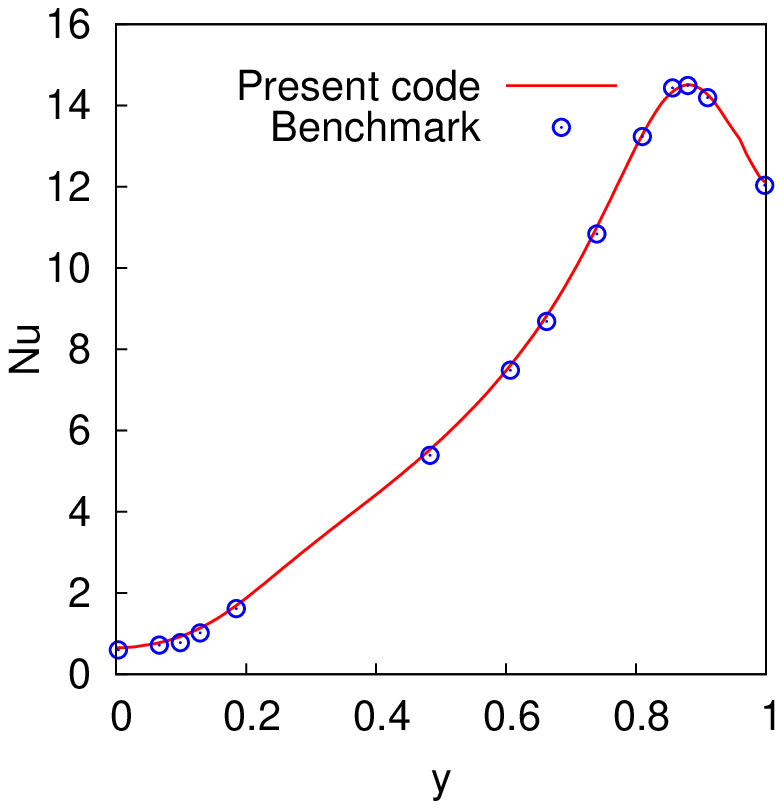}
  \caption{Distribution of the local Nusselt number along the vertical cold wall compared to the benchmark data of Demirdzic et al. \cite{demir1992}. \label{val_Nu}}
\end{minipage} 
\end{figure}

Here we use the benchmark test case proposed by Demirdzic et al. \cite{demir1992} to validate the implementation of natural convection  in the code. A cylinder is placed inside a square enclosure of dimensions $L\times L$ with $L=1$. The cylinder has a diameter of $d/L=0.4$ and its center is displaced vertically by $\delta/L = 0.1$. The cylinder is heated at $T_h = 1$, while the vertical walls of the duct are cooled at $T_c=0$ and the horizontal walls are assumed to be adiabatic. The Rayleigh number based on $L$ is $10^6$, and the Prandtl number is $0.1$. The benchmark solution of Demirdzic et al. \cite{demir1992}  was obtained using a structured mesh of $256\times 128$. Because of the symmetry of the problem, only one-half of the domain needs to be modeled. For our computation we used a mesh composed of $15,366$ cells, with quadrilateral cells around the walls and triangular cells elsewhere as showed in Fig. \ref{valid_convnat}. In this figure are presented the calculated streamlines and temperature contours that compare well  with those of \cite{demir1992}. The local Nusselt number along the vertical wall of the duct is presented in Fig. \ref{val_Nu}. A very good agreement is found between our results and those of the benchmark. Furthermore, the calculated mean Nusselt number at this wall is $6.7195$ while the benchmark value is $6.7303$ which gives a relative difference of less than $0.1\ \%$. 

%\newpage
%\listoffigures
%\newpage
%\listoftables
\end{document}